\documentclass[12pt,pdf,epsfig,color,png]{article}
\usepackage{amsmath}
\usepackage{amsfonts}
\usepackage{amsmath}
\usepackage{slashed}
\usepackage{amsxtra}
\usepackage{amstext}
\usepackage{amssymb}
\usepackage{color}
\usepackage{float}
\usepackage{graphicx}
\usepackage{subcaption}
\numberwithin{equation}{section}
\usepackage[dvips]{epsfig}
\usepackage{appendix}
\usepackage{youngtab}
\allowdisplaybreaks

\newcommand{\be}{\begin{equation}}
\newcommand{\ee}{\end{equation}}
\newcommand{\beq}{\begin{equation}}
\newcommand{\eeq}{\end{equation}}
\newcommand{\ba}{\begin{eqnarray}}
\newcommand{\ea}{\end{eqnarray}}
\newcommand{\bea}{\begin{eqnarray}}
\newcommand{\eea}{\end{eqnarray}}

\begin{document}
\baselineskip=15.5pt \pagestyle{plain} \setcounter{page}{1}
%
\begin{titlepage}

\vskip 0.8cm

\begin{center}
%
%
%
%
%

{\Large \bf Four-dilatino superstring theory scattering amplitude and its Regge behavior}

\vskip 1.cm

{\large {{\bf Lucas Martin{\footnote{\tt lucasmartinar@iflp.unlp.edu.ar}}, Martin Parlanti{\footnote{\tt martin.parlanti@fisica.unlp.edu.ar}}} {\bf and  Martin
Schvellinger}{\footnote{\tt martin@fisica.unlp.edu.ar}}}}

\vskip 1.cm

{\it Instituto de F\'{\i}sica La Plata-UNLP-CONICET. \\
Boulevard 113 e 63 y 64, (1900) La Plata, Buenos Aires, Argentina \\
and \\
Departamento de F\'{\i}sica, Facultad de Ciencias Exactas,
Universidad Nacional de La Plata. \\
Calle 49 y 115, C.C. 67, (1900) La Plata, Buenos Aires, Argentina.}

\vspace{1.cm}

{\bf Abstract}

\end{center}

\vspace{.5cm}

An explicit derivation of the four-dilatino scattering amplitude from type II superstring theory is presented. This is obtained by using the Kawai-Lewellen-Tye relations for closed string scattering amplitudes in terms of the product of two open string scattering amplitudes. We also investigate its Regge limit.

\noindent

\end{titlepage}

\newpage

{\small \tableofcontents}

\newpage

%
%
\section{Introduction}
%
%

In 1985 Kawai, Lewellen and Tye (KLT) proposed a general formula to calculate any tree-level closed string scattering amplitude as a sum of products of certain open string scattering amplitudes at tree level \cite{Kawai:1985xq}. The purpose of the present work is to give an explicit realization of the KLT relation for the specific case of four dilatinos in type II superstring theory. We hope this will be useful for further calculations and applications. For instance, the results presented in this work are useful in the context of the holographic dual calculation of hard scattering of fermions \cite{Martin:2024jpe}, including the Regge behavior and the fixed angle limit.

The historical development of the string theory scattering amplitudes is very interesting. It can be traced back to 1968 when, in a ground-breaking work \cite{Veneziano:1968yb} Veneziano proposed the construction of a crossing-symmetric scattering amplitude of four particles with Regge behavior. Soon after, Virasoro developed an alternative construction of that idea \cite{Virasoro:1969me}, which Shapiro extended to $N$ particles \cite{Shapiro:1970gy}. Later on, it was realized that the amplitudes presented in \cite{Veneziano:1968yb} and \cite{Virasoro:1969me} can be straightforwardly derived from scattering of four open strings and four closed strings, respectively. Veneziano's amplitude, initially associated with four-meson scattering in the old hadronic string theory, actually represents the four-tachyon scattering amplitude of open bosonic string theory. Virasoro's amplitude, on the other hand, corresponds to four-closed strings scattering. The Ramond-Neveu-Schwarz (RNS) formalism leading to worldsheet supersymmetry was proposed in \cite{Ramond:1971gb,Neveu:1971rx}. The next step was the construction of a formalism considering spacetime supersymmetry in string theory, which was developed by Green and Schwarz \cite{Green:1981yb,Green:1981xx}.

In this work we focus on the closed-string scattering amplitude of four spin-1/2 fermions in type II superstring theory at tree level. The construction of the fermion-emission vertex and the calculation of fermion-fermion scattering amplitudes for open strings were developed in several papers. In particular, the fermion emission vertex operator was developed in references \cite{Thorn:1971jc,Schwarz:1971uie,Corrigan:1972tg}. Gauge conditions for the dual fermion emission vertex were studied in \cite{Brink:1973jd}, while reference \cite{Olive:1973ewy} focused on the construction of scattering amplitudes with four external fermions (with a high degree of internal consistency based on Lorentz invariance and the no-ghost theorem) although the authors could not evaluate explicitly the amplitudes. Mandelstam \cite{Mandelstam:1973je} extended the interacting-string formalism to the Neveu-Schwarz and Ramond models and pointed out that dual open-string scattering amplitudes including any number of external fermions can be constructed. In addition, that paper showed that these fermionic scattering amplitudes are manifestly free of ghosts. Furthermore, Schwarz and Wu evaluated amplitudes for scattering of fermion ground states in the context of dual resonance models for fermions and mesons in the operator formalism \cite{Schwarz:1973jf,Schwarz:1974ie}. Corrigan, Goddard, Smith and Olive \cite{Corrigan:1973tye} calculated the scattering amplitude of four fermions within the operator formalism and showed that their results agree with Mandelstam's results \cite{Mandelstam:1973je}, where the same amplitudes were derived using the open string formalism \cite{Goddard:1973qh,Mandelstam:1973jk}. The amplitudes obtained in \cite{Mandelstam:1973je,Corrigan:1973tye} are not quite invariant under crossing symmetry \cite{Corrigan:1973tye}.

As pointed out by Schwarz \cite{Schwarz:1982jn} the calculation of the fermion-fermion scattering amplitude in the RNS formalism carried out during the early 70's demanded an enormous work. Although the light-cone gauge quantization was very important for the initial formulation of the theory and also for certain calculations \cite{Goddard:1973qh}, the covariant quantization of superstring theories led to a deeper understanding. The covariant quantization was developed within the framework of conformal field theory and the BRST quantization \cite{Friedan:1985ge}. The covariant fermionic vertex operator in the RNS string model was constructed in \cite{Knizhnik:1985ke}. The calculation of the four-fermion open-string scattering amplitude in the framework of the RNS model was developed in \cite{Cohn:1986bn} using the covariant quantization, leading to an explicit short expression (also presented in \cite{Green:1987sp}). Furthermore, covariant fermionic scattering amplitudes were obtained by Knizhnik \cite{Knizhnik:1986ke}. In addition, three- and four-point scattering amplitudes of massless string states 
of the heterotic string were calculated by Gross, Harvey, Martinec and Rohm \cite{Gross:1984dd,Gross:1985fr,Gross:1985rr}.

At this point it is interesting to comment on the differences between our present work with respect to the one by Atick and Sen \cite{Atick:1986rs}. They used the RNS formulation to calculate the two fermion-to-two boson and the four-fermion closed-string scattering amplitudes at one loop of the heterotic and type II superstring theories using the covariant fermion emission vertex. Moreover, their results agree with the amplitudes calculated in the light-cone gauge in the Green-Schwarz formulation. However, it is important to mention that their result in the case of four fermions (given by their equation (3.68)) is expressed in terms of the product of the kinematic factor of the four-fermion scattering amplitude for open strings which they called $K(u_{(1)}, u_{(2)}, u_{(3)}, u_{(4)}; k_i)$, given by their equation (3.67), times an integral which depends on the properly normalized contribution from the anti-analytic sector. Thus, they did not give an explicit expression of the four-fermion scattering amplitude. We should emphazise that although in reference \cite{Atick:1986rs} it was considered a one-loop calculation of several closed-string scattering amplitudes, even at tree level the main difference with our calculation is that we have used the KLT relation and worked in the Green-Schwarz formalism. Moreover, we have obtained the explicit form of the four spin-1/2 fermions closed-string scattering amplitude.\footnote{The four-point open-string scattering amplitudes which are relevant for the present work are presented in \cite{Green:1981xx,Schwarz:1982jn,Becker:2015eia}. Let us recall that, using the supersymmetric formulation in ten dimensions, Green and Schwarz constructed vertex operators for the emission of massless open-strings states \cite{Green:1981yb,Green:1981xx} in the light-cone gauge formalism which they previoulsy developed in reference \cite{Green:1980zg}. They also constructed various four-particle superstring theory scattering amplitudes for massless vectors and spinors, including one-loop calculations of the scattering amplitudes \cite{Green:1981ya,Green:1982sw}.} Also, we have considered the full kinematics leading to an explicit expression in terms of the Mandelstam variables and the scattering angle. In addition, we have obtained the Regge limit of four-closed string scattering amplitude of type II superstring theory involving four spin-1/2 fermions.

It is worth mentioning that in reference \cite{Becker:2015eia} some tree-level three-, four- and five-point string theory scattering amplitudes were studied in the RNS formalism for open strings. Then, by using the KLT relations they investigated some closed string scattering amplitudes including two and four gravitinos. However, for instance their expressions (6.22)  and (6.23) with two NS-NS fields and two NS-R gravitinos do not give the explicit forms of these scattering amplitudes, which is effectively a very laborious calculation. Dilatinos were not considered in that reference.

In addition, during the last 25 years the pure spinor formulation for the superstring has been developed after the pioneering work of Berkovits \cite{Berkovits:2000fe}. In \cite{Berkovits:2000ph} it was proven that for the scattering of an arbitrary number of massless bosons and up to four massless fermions, tree-level scattering amplitudes for open strings agree with the ones obtained by using the RNS formalism \cite{Friedan:1985ge}. The equivalence with the RNS formalism was further extended also at tree level for type I string theory scattering amplitudes involving up to four fermions \cite{Alencar:2011tk}. The pure spinor formalism somehow simplifies the calculation of multiparticle and multiloop scattering amplitudes. Also, it can be applied beyond the flat ten-dimensional spacetime, allowing for the description of Ramond-Ramond backgrounds. For a recent review see \cite{Berkovits:2022fth} and references therein. For a comprehensive review of recent developments of the pure spinor formalism deriving massless superstring scattering amplitudes at tree level we refer the reader to the report by Mafra and Schlotterer \cite{Mafra:2022wml}.

There are also important extensions of the KLT relations. For instance, one-loop string theory scattering amplitudes involving both open and closed strings were expressed as a sum over pure open string amplitudes \cite{Stieberger:2021daa}. Furthermore, a relation between one-loop amplitudes of closed and open strings which generalizes the KLT relation to one-loop string theory scattering amplitudes was proposed by Stieberger  \cite{Stieberger:2022lss}.

~

This work is organized as follows. In section 2, as an example of the use of the KLT relations we explicitly derive the well-known four-dilaton scattering amplitude. Our main result, which is the explicit derivation of the four-dilatino superstring scattering amplitude in type II superstring theory, is presented in section 3. In section 4 we study the high energy behavior, obtaining the Regge limit of the scattering amplitude. Conclusions are given in section 5. We also present more details of the calculations in two appendices.

%
%
\section{A derivation of the four-dilaton scattering amplitude}
%
%

In this section we present a review of the calculation of the four-boson closed superstring scattering amplitude, which is a very instructive calculation before addressing the main topic of this work in section 3. The kinematic part of the four-dilaton scattering amplitude, which corresponds to four closed strings in type II superstring theory, is given by 
\begin{equation}
K_{\text{closed}}^{\text{bosonic}}(1,2,3,4)= \frac{1}{16}K_{\text{open}}^{\text{bosonic}}(1,2,3,4)\otimes \frac{1}{16} K_{\text{open}}^{\text{bosonic}}(1,2,4,3) \, , \label{Kclosed4dilatons}
\end{equation}
where factors $1/16$ come from the fact that
\begin{equation}
\frac{1}{16} K_{\text{open}}^{\text{bosonic}}(1,2,3,4)\equiv\frac{1}{16} K_{\text{open}}^{\text{bosonic}}(k_1, k_2, k_3, k_4) = K_{\text{open}}^{\text{bosonic}}(k_1/2, k_2/2, k_3/2, k_4/2) \, , \label{Kopen4dilatons}
\end{equation}
i.e. in the open string kinematic factor one must replace the ten-dimensional momenta $k_i^M$ corresponding to each closed string by half of them, as required by the KLT relations. $K_{\text{open}}^{\text{bosonic}}$ is invariant under $S_4$ permutation symmetry.  Notice that in the second factor of equation ({\ref{Kclosed4dilatons}}) the strings 3 and 4 are exchanged in comparison with the first factor. This is demanded by the KLT relations.

The $K_{\text{open}}^{\text{bosonic}}(1,2,3,4)$ factor \cite{Green:1981xx,Schwarz:1982jn,Green:1987sp} can be written as
\begin{eqnarray}
{K_{\text{open}}^{\text{bosonic}}}(1,2,3,4)&=&-\frac{1}{4}\left(\tilde{s} \ \tilde{u} \ \zeta_1\cdot\zeta_4 \ \zeta_3\cdot \zeta_2+\tilde{s} \ \tilde{t} \ \zeta_1\cdot\zeta_3 \ \zeta_4\cdot \zeta_2+\tilde{t} \ \tilde{u} \ \zeta_1\cdot\zeta_2 \ \zeta_4\cdot \zeta_3\right)\nonumber\\
&+&\frac{\tilde{s}}{2}\left(\zeta_1\cdot k_4 \ \zeta_3\cdot k_2 \ \zeta_4\cdot \zeta_2 +\zeta_2\cdot k_3 \ \zeta_4\cdot k_1\zeta_1\cdot \zeta_3 +\zeta_1\cdot k_3 \ \zeta_4\cdot k_2 \ \zeta_3\cdot \zeta_2 \right. \nonumber \\
&& \left. + \ \zeta_2\cdot k_4 \ \zeta_3\cdot k_1 \ \zeta_1\cdot \zeta_4 \right) \nonumber\\
&+& \frac{\tilde{t}}{2} \left(\zeta_2\cdot k_1 \ \zeta_4\cdot k_3\zeta_3\cdot \ \zeta_1+\zeta_3\cdot k_4 \ \zeta_1\cdot k_2 \ \zeta_2\cdot \zeta_4+\zeta_2\cdot k_4 \ \zeta_1\cdot k_3 \ \zeta_3\cdot \zeta_4 \right. \nonumber \\
&& \left. + \ \zeta_3\cdot k_1 \ \zeta_4\cdot k_2 \ \zeta_2\cdot \zeta_1\right)\nonumber\\
&+& \frac{\tilde{u}}{2} \left(\zeta_1\cdot k_2 \ \zeta_4\cdot k_3 \ \zeta_3\cdot \zeta_2+\zeta_3\cdot k_4 \ \zeta_2\cdot k_1 \ \zeta_1\cdot \zeta_4+\zeta_1\cdot k_4 \ \zeta_2\cdot k_3 \ \zeta_3\cdot \zeta_4 \right. \nonumber \\
&& \left. +\zeta_3\cdot k_2 \ \zeta_4\cdot k_1 \ \zeta_1\cdot \zeta_2 \right) \, , 
\label{Kopen4bosons}
\end{eqnarray}
where the ten-dimensional Mandelstam variables are defined as
\begin{eqnarray}
\tilde{s}&=&-(k_1+k_2)^2 \, , \nonumber\\
\tilde{t}&=&-(k_1+k_4)^2 \, , \nonumber\\
\tilde{u}&=&-(k_1+k_3)^2 \, , \label{Mandelstam-var}
\end{eqnarray}
with the convention that all the momenta are incoming. The Mandelstam variables satisfy $\tilde{s}+\tilde{t}+\tilde{u}=0$ since we only consider massless string states. We use the mostly plus Minkowski ten-dimensional metric $\eta_{MN} \equiv \text{diag}(-1, 1, \dots, 1)$, where $M, N, \cdots$ are ten-dimensional Lorentz indices.

Plugging equation (\ref{Kopen4bosons}) into equation (\ref{Kclosed4dilatons}) there are in principle 225 terms from which only 120 are independent. They involve the product of two vector polarizations of the form $\zeta_i^M\otimes \zeta_i^{M'}$, corresponding to the left-moving and right-moving pieces of the closed-string scattering polarization tensor $\Theta_i^{MM'}$. Notice that the subindex $i$ indicates each of the four dilatons involved in the scattering process, $\Phi_i$. Thus, one must replace the product of the two polarization vectors by a rank-2 tensor as follows \cite{Gross:1986mw}
\begin{eqnarray}
\zeta_i^M\otimes \zeta_i^{M'} \rightarrow \Theta_i^{MM'} \, .
\end{eqnarray}
The transverse diagonal part corresponds to the dilaton polarization 
\begin{eqnarray}
\Theta_i^{MM'}=\frac{1}{\sqrt{8}} \left(\eta^{MM'}-k_i^M\bar{k}_i^{M'}-k_i^{M'}\bar{k}_i^{M}\right)  \Phi_i \, , \label{rank-2-tensor}
\end{eqnarray}
with the conditions $\bar{k} \cdot \bar{k}=0$ and $k \cdot \bar{k}=1$. By using this dilaton polarization tensor each of the 120 terms mentioned before contains the product of four factors of the form (\ref{rank-2-tensor}). Thus, it leads to 9720 terms in total contributing to the four-dilaton scattering kinematic factor, which after some simple but very tedious algebra gives 
\begin{eqnarray}
K_{\text{closed}}^{\text{4-dilatons}}(1,2,3,4)&=&\frac{1}{128} \left[6 \tilde{s}^4-15\tilde{s}^3(\tilde{t}+\tilde{u})+\tilde{s}^2(30 \tilde{t}^2+28\tilde{t}\tilde{u}+30 \tilde{u}^2) \right. \nonumber \\
&& \left. -\tilde{s}(\tilde{t}+\tilde{u})(15 \tilde{t}^2-43\tilde{t}\tilde{u}+15\tilde{u}^2) \right. \nonumber \\
&& \left. + 3(2\tilde{t}^4-5\tilde{t}^3\tilde{u}+10\tilde{t}^2\tilde{u}^2-5\tilde{t}\tilde{u}^3+2\tilde{u}^4)\right] 
\Phi_1 \Phi_2 \Phi_3 \Phi_4 \, .
\end{eqnarray}
Since $\tilde{u}=-\tilde{s}-\tilde{t}$, the kinematical factor $K_{\text{closed}}^{\text{4-dilatons}}(1,2,3,4)$ becomes
\begin{eqnarray}
K_{\text{closed}}^{\text{4-dilatons}}(1,2,3,4)&=& \frac{9}{16}\left(\tilde{s}^2+\tilde{s} \ \tilde{t}+\tilde{t}^2\right)^2 \Phi_1 \Phi_2 \Phi_3 \Phi_4 \, .
\end{eqnarray}
The Mandelstam variables $\tilde{t}$ and $\tilde{u}$ can be rewritten in terms of the scattering angle $\theta$
\begin{eqnarray}
\tilde{t}&=&-\frac{\tilde{s}}{2}(1-\cos\theta) \, , \nonumber \\ 
\tilde{u}&=&-\frac{\tilde{s}}{2}(1+\cos\theta) \, .
\end{eqnarray}
Then, the kinematic factor can be expressed as\footnote{The same result is obtained from the expression $(9/32) (\tilde{s}^4+\tilde{t}^4+\tilde{u}^4) \Phi_1 \Phi_2 \Phi_3 \Phi_4$. This is
known as the kinematic factor associated with the Virasoro-Shapiro-Green-Schwarz-Brink scattering amplitude for closed strings \cite{Virasoro:1969me,Shapiro:1970gy,Green:1982sw}.}
\begin{equation}
K_{\text{closed}}^{\text{4-dilatons}}(1,2,3,4)=\frac{9 \tilde{s}^4 [\cos (2 \theta)+7]^2}{1024} \Phi_1 \Phi_2 \Phi_3 \Phi_4 \, .
\end{equation}
Therefore, the four-dilaton closed-string scattering amplitude is given by 
\begin{eqnarray}
&&\mathcal{A}_{\text{closed}}^{\text{4-dilatons}}(k_1,k_2,k_3,k_4) =  -4 \pi g_s^2 \alpha'^7 \left(\prod_{\chi=\tilde{s}, \tilde{t}, \tilde{u}} \frac{\Gamma(-\alpha' \chi/4)}{\Gamma(1+\alpha' \chi/4)}\right) \ K_{\text{closed}}^{\text{4-dilatons}}(1,2,3,4) \nonumber \\
&=&-4\pi g_s^2 \alpha'^3 \frac{9 (\alpha'\tilde{s})^4 [\cos (2 \theta)+7]^2}{1024} \left(\prod_{\chi=\tilde{s},\tilde{t},\tilde{u}} \frac{\Gamma(-\alpha^\prime \chi/4)}{\Gamma(1+\alpha^\prime \chi/4)} \right) \Phi_1 \Phi_2 \Phi_3 \Phi_4 \, , \nonumber \\
&&
\end{eqnarray}
where we have written the Euler gamma functions, while $\alpha'$ and $g_s$ are the string constant and string coupling, respectively.

Next, we can consider the Regge limit ($\tilde{s} \gg |\tilde{t}|$) of this scattering amplitude leading to:
\begin{eqnarray}
&&\mathcal{A}_{\text{closed}}^{\text{4-dilatons Regge}}(k_1,k_2,k_3,k_4) =  4 \pi g_s^2 \alpha'^7\ K_{\text{closed}}^{\text{4-dilatons Regge}}(1,2,3,4) \times \nonumber \\
&&\frac{\sin[\frac{\pi\alpha'}{4}(\tilde{s}+\tilde{t})]}{\sin(\frac{\pi\alpha'\tilde{s}}{4})} \left(\frac{\alpha'\tilde{s}}{4}\right)^{-2+\frac{\alpha'\tilde{t}}{2}} \ e^{2-\frac{\alpha'\tilde{t}}{2}}\frac{\Gamma(-\alpha'\tilde{t}/4)}{\Gamma(1+\alpha'\tilde{t}/4)} \nonumber \\
&=&36\pi g_s^2 \alpha'^3 \, 2^{-\alpha'\tilde{t}} \, \, \frac{\sin[\frac{\pi\alpha'}{4}(\tilde{s}+\tilde{t})]}{\sin(\frac{\pi\alpha'\tilde{s}}{4})} \left(\alpha'\tilde{s}\right)^{2+\frac{\alpha'\tilde{t}}{2}} e^{2-\frac{\alpha'\tilde{t}}{2}}\frac{\Gamma(-\alpha'\tilde{t}/4)}{\Gamma(1+\alpha'\tilde{t}/4)} \Phi_1 \Phi_2 \Phi_3 \Phi_4 \, , \label{Four-dilaton-Regge}
\end{eqnarray}
where we have used the Stirling's formula to approximate the Euler gamma functions, which only holds in the Regge limit.

%
%
\section{Derivation of the four-dilatino scattering amplitude}
%
%

In this secion we derive the four-dilatino scattering amplitude from type II superstring theory.  We explicitly show how to  use the Kawai-Lewellen-Tye relations for closed string amplitudes in terms of the product of two open-string scattering amplitudes.

The kinematic part of the amplitude of four-dilatino scattering amplitude, which corresponds to four closed strings in type II superstring theory, is given by 
\begin{equation}
K_{\text{closed}}^{\text{4-dilatino}}(\tilde{1},\tilde{2}, \tilde{3},\tilde{4})= K_{\text{open}}^{\text{fermionic}}(\tilde{1},\tilde{2}, \tilde{3},\tilde{4})\otimes \frac{1}{16} K_{\text{open}}^{\text{bosonic}}(1,2,4,3) \, ,\label{Kclosed4dilatinos} 
\end{equation}
where factor $K_{\text{open}}^{\text{bosonic}}(1,2,4,3)$ is given in equation (\ref{Kopen4bosons}), and the Mandelstam variables have been defined in equation (\ref{Mandelstam-var}).
As in the bosonic case described in section 2, we only consider massless closed string states, therefore we have $\tilde{s}+\tilde{t}+\tilde{u}=0$. In the second factor of equation ({\ref{Kclosed4dilatinos}}) the strings 3 and 4 are exchanged, as required by the KLT relations. The order of the two kinematic factors in ({\ref{Kclosed4dilatinos}}) is due to the fact that we calculate it for R-NS dilatinos. In addition, it is interesting to mention that from this equation the four-gravitino scattering amplitude could be obtained.

On the other hand, the kinematic factor $K_{\text{open}}^{\text{fermionic}}(\tilde{1},\tilde{2}, \tilde{3},\tilde{4})$ is given by \cite{Green:1981xx,Schwarz:1982jn,Green:1987sp} 
\begin{eqnarray}
K_{\text{open}}^{\text{fermionic}}(\tilde{1},\tilde{2},\tilde{3},\tilde{4} )&=& -\frac{1}{8} \tilde{s} \bar{u}_2 \Gamma^{M'} u_{3} \bar{u}_1\Gamma_{M'} u_4 +  \frac{1}{8} \tilde{t} \bar{u}_1 \Gamma^{M'} u_{2} \bar{u}_4  \Gamma_{M'} u_3 \, , \label{Kopen4fermions}
\end{eqnarray} 
where $u_i$ represents the spinor polarization, while 
the ten-dimensional Gamma matrices satisfy the Clifford algebra
\begin{eqnarray}
 \lbrace\Gamma^M,\Gamma^N\rbrace=-2\eta^{MN}   \, .
\end{eqnarray}
The ten-dimensional Gamma matrices have $32\times32$ pure imaginary components since they are in the Majorana representation given by \cite{Green:1987sp}:
\begin{eqnarray}
    \Gamma^0&=&\sigma^2 \otimes 1_{16}\\
    \Gamma^j&=&i\sigma^1 \otimes \gamma^j, j=1,..,8 \\
    \Gamma^9&=&i\sigma^3 \otimes 1_{16} \, ,
\end{eqnarray}
where $\gamma^j$'s are the 16-dimensional matrices corresponding to the reducible ${\bf {8_s}}+{\bf {8_c}}$ representation of spin(8), which can be written as\footnote{The labels $16 \times 16$ and then $8 \times 8$ make explicitly the dimension of these matrices.}
\begin{eqnarray}
\gamma^j=\begin{pmatrix}0 & \gamma^j_{a \dot{a}} \\  \gamma^j_{\dot{b} b} & 0 
\end{pmatrix}_{16 \times 16}, 
\end{eqnarray}
being  $\gamma^j_{\dot{a} a} = (\gamma^j_{a \dot{a}})^t$. These $\gamma^j$ matrices satisfy the Clifford algebra if the following condition is fulfilled:
$\gamma^i_{a \dot{a}} \gamma^j_{\dot{a} b} + \gamma^j_{a \dot{a}} \gamma^i_{\dot{a} b} =  2 \delta^{ij} \delta_{ab}$,
and a similar algebraic condition with the exchange of dotted and undotted indices. In particular, a specific set of eight-dimensional $\gamma^j_{a \dot{a}} \rightarrow \gamma^j_{8 \times 8}$ matrices which satisfies these conditions can be constructed in the following way:
%
\begin{eqnarray}
&& \gamma^1_{8 \times 8} = \epsilon \times \epsilon \times \epsilon \, , \,\,\,\,\,\,\,\,\,\,\,\,\,\,\,\,\,\,\,\, \gamma^2_{8 \times 8} = I_2 \times \sigma^1 \times \epsilon \, , \nonumber \\
&& \gamma^3_{8 \times 8} = I_2 \times \sigma^3 \times \epsilon \, , \,\,\,\,\,\,\,\,\,\,\,\,\,\, \gamma^4_{8 \times 8} = \sigma^1 \times \epsilon \times I_2 \, , \nonumber \\
&& \gamma^5_{8 \times 8} = \sigma^3 \times \epsilon \times 1_2 \, , \,\,\,\,\,\,\,\,\,\,\,\,\,\, \gamma^6_{8 \times 8} = \epsilon \times I_2 \times \sigma^1 \, , \nonumber \\
&& \gamma^7_{8 \times 8} = \epsilon \times 1_2 \times \sigma^3 \, , \,\,\,\,\,\,\,\,\,\,\,\,\,\, \gamma^8_{8 \times 8} = I_2 \times I_2 \times I_2 \, , 
\end{eqnarray}
where the definition $\epsilon \equiv i \sigma^2$ and the $d \times d$ identity matrix $I_d \equiv I_{d \times d}$ have been used. The Pauli matrices are defined as:
\begin{eqnarray}
\sigma^1=\begin{pmatrix}0 & 1 \\ 1 & 0 
\end{pmatrix}, \,\,\,\,\,
\sigma^2=\begin{pmatrix}0 & -i \\ i & 0\end{pmatrix},\,\,\,\,\,
\sigma^3=\begin{pmatrix}1 & 0\\ 0& -1 
\end{pmatrix}\, .
\end{eqnarray}

Plugging equations (\ref{Kopen4fermions}) and (\ref{Kopen4bosons}) into equation (\ref{Kclosed4dilatinos}) one obtains 30 terms. These terms contain the direct product of a fermion polarization and a vector polarization of the generic form $u_i^\alpha\otimes \zeta_i^M$. This allows us to define  the ten-dimensional dilatino $\lambda_i^\beta$ as: 
\begin{eqnarray}
u_i^\alpha\otimes \zeta_i^M =i(\Gamma^M)^\alpha_\beta \lambda_i^\beta    \, ,
\end{eqnarray}
where the fermionic indices $\alpha$ and $\beta$ run from 1 to 32, while the sub-index $i$ denotes each dilatino \cite{Garousi:1996ad}. Also notice that $u_i^\alpha$ is the $\alpha$ component of the spinor polarization $u_i$ from the open-string kinematic factor (\ref{Kopen4fermions}). The imaginary unit in the definition of the fermion polarization can be removed since it has no effect on our calculation of the four-fermion scattering amplitude.

Now, we proceed to carry out the explicit calculation of the tensor product (\ref{Kclosed4dilatinos}). We may write it in a concise form:
\begin{equation}
K_{\text{closed}}^{\text{4-dilatino}}(\tilde{1},\tilde{2}, \tilde{3},\tilde{4})= \sum_{j=1}^{15} ({\cal {T}}_{1,j}+{\cal{T}}_{2,j}) \, ,\label{SumKclosed4dilatinos} 
\end{equation}
where ${\cal {T}}_{1,j}$ is given by the tensor product of the first term of equation (\ref{Kopen4fermions}) with the $j$-th term of equation (\ref{Kopen4bosons}). In total, this leads to the first 15 terms of $K_{\text{closed}}^{\text{4-dilatino}}(\tilde{1},\tilde{2}, \tilde{3},\tilde{4})$ in equation (\ref{Kclosed4dilatinos}). Similarly, we have ${\cal {T}}_{2,j}$ which represents each of the 15 terms given by the tensor products of the second term of equation (\ref{Kopen4fermions}) with the $j$-th term of equation (\ref{Kopen4bosons}). We should mention that in both cases ${\cal {T}}_{i,j}$'s are defined considering the exchange of the particles 3 and 4 in equation (\ref{Kopen4bosons}) as required by the KLT relations. Below we present two examples. The first one is
\begin{eqnarray}
    {\cal {T}}_{1,1}&=&\left(-\frac{1}{8} \tilde{s} \bar{u}_2 \Gamma^{M'} u_{3} \bar{u}_1\Gamma_{M'} u_4\right) \otimes \left(-\frac{1}{16\cdot 4}\tilde{s}\tilde{t}\zeta_1^M \zeta_{3,M} \zeta_4^Q \zeta_{2,Q}\right)\nonumber\\
    &=& \frac{\tilde{s}^2\tilde{t}}{512}\bar{\lambda}_2\Gamma_N\Gamma_P\Gamma_M\lambda_3\bar{\lambda}_1\Gamma^M\Gamma^P\Gamma^M\lambda_4 \, ,
\end{eqnarray}
and the second tensor product is
\begin{eqnarray}
     {\cal {T}}_{1,2}&=&\left(-\frac{1}{8} \tilde{s} \bar{u}_2 \Gamma^{M'} u_{3} \bar{u}_1\Gamma_{M'} u_4\right)\otimes  \left(-\frac{1}{16\cdot 4}\tilde{s}\tilde{u}\zeta_1^M \zeta_{4,M} \zeta_3^P \zeta_{2,P}  \right) \nonumber\\
    &=&\frac{1}{8}\tilde{s}^2\tilde{u}\bar{\lambda_2}\Gamma^M\lambda_3\bar{\lambda_1}\Gamma_M\lambda_4 \, .
\end{eqnarray}
In order for the interested reader to be able to reproduce the details of the calculations, the full list of the 30 terms is presented in appendix A.

The result of adding the 30 terms of appendix A as required by equation (\ref{SumKclosed4dilatinos}) gives
\begin{eqnarray}
 &  & K_{\text{closed}}^{\text{4-dilatino}}(\tilde{1},\tilde{2},\tilde{3},\tilde{4} )= \nonumber \\
&& \frac{\tilde{s}^2\tilde{t}}{512}\bar{\lambda}_2\Gamma_N\Gamma_P\Gamma_M\lambda_3\bar{\lambda}_1\Gamma^M\Gamma^P\Gamma^N\lambda_4-\frac{\tilde{s}\tilde{t}^2}{512}\bar{\lambda_1}\Gamma^M\Gamma^P\Gamma^N\lambda_2\bar{\lambda_4}\Gamma_N\Gamma_P\Gamma_M\lambda_3 \nonumber \nonumber \\
&+&\frac{1}{8}\tilde{s}^2\tilde{u}\bar{\lambda_2}\Gamma^M\lambda_3\bar{\lambda_1}\Gamma_M\lambda_4
-\frac{1}{8}\tilde{t}^2\tilde{u}\bar{\lambda_1}\Gamma^M\lambda_2\bar{\lambda_4}\Gamma_M\lambda_3+\frac{\tilde{s}\tilde{t}\tilde{u}}{512}\bar{\lambda}_2\Gamma_M\Gamma_P\Gamma_N\lambda_3\bar{\lambda}_1\Gamma^M\Gamma^P\Gamma^N\lambda_4 \nonumber \\
&-&\frac{\tilde{s}\tilde{t}\tilde{u}}{512}\bar{\lambda_1}\Gamma^M\Gamma^N\Gamma^P\lambda_2\bar{\lambda_4}\Gamma_M\Gamma_N\Gamma_P\lambda_3 -\frac{\tilde{s}^2}{32}\bar{\lambda}_2\Gamma_M\lambda_3\bar{\lambda}_1(k_3\cdot \Gamma)\Gamma^M(k_2\cdot\Gamma)\lambda_4 \nonumber \\
&+&\frac{\tilde{t}^2}{32}\bar{\lambda}_1(k_3 \cdot \Gamma)\Gamma^M (k_4 \cdot \Gamma)\lambda_2\bar{\lambda}_4\Gamma_M \lambda_3 -\frac{\tilde{s}^2}{32}\bar{\lambda}_2(k_4 \cdot \Gamma)\Gamma^M(k_1 \cdot \Gamma)\lambda_3\bar{\lambda}_1\Gamma_M\lambda_4  \nonumber \\
&+& \frac{\tilde{t}^2}{32}\bar{\lambda}_1 \Gamma_M\lambda_2\bar{\lambda}_4(k_2 \cdot \Gamma)\Gamma^M (k_1 \cdot \Gamma)\lambda_3 -\frac{\tilde{s}^2}{16}k_2^{[M}\bar{\lambda}_2\Gamma^{N]}\lambda_3 k_{4[M}\bar{\lambda}_1\Gamma_{N]}\lambda_4 \nonumber \\
&+& \frac{\tilde{t}^2}{16}k_2^{[N}\bar{\lambda}_1\Gamma^{M]}\lambda_2 k_{4[N}\bar{\lambda}_4\Gamma_{M]}\lambda_3-\frac{\tilde{s}^2}{16}k_{3[M}\bar{\lambda}_2\Gamma_{N]}\lambda_3 k_1^{[M}\bar{\lambda}_1\Gamma^{N]}\lambda_4 +\frac{\tilde{t}^2}{16}k_1^{[N}\bar{\lambda}_1\Gamma^{M]}\lambda_2 k_{3[N}\bar{\lambda}_4\Gamma_{M]}\lambda_3 \nonumber \\
&-&\frac{\tilde{s}\tilde{u}}{32}\bar{\lambda}_2(k_1 \cdot \Gamma)\Gamma^M(k_4 \cdot \Gamma)\lambda_3\bar{\lambda}_1\Gamma_M\lambda_4 +\frac{\tilde{t}\tilde{u}}{32}\bar{\lambda}_1 \Gamma_M\lambda_2\bar{\lambda}_4(k_1 \cdot \Gamma)\Gamma^M (k_2 \cdot \Gamma)\lambda_3 \nonumber \\
&-&\frac{\tilde{s}\tilde{u}}{32}\bar{\lambda}_2\Gamma_M\lambda_3\bar{\lambda}_1(k_2 \cdot \Gamma)\Gamma^M(k_3 \cdot \Gamma)\lambda_4+\frac{\tilde{t}\tilde{u}}{32}\bar{\lambda}_1(k_4 \cdot \Gamma)\Gamma^M (k_3 \cdot \Gamma)\lambda_2\bar{\lambda}_4\Gamma_M \lambda_3\nonumber\\
&-&\frac{\tilde{s}\tilde{u}}{16}k_3^{[N}\bar{\lambda}_2 \Gamma^{M]}\lambda_3k_{4[N}\bar{\lambda}_1\Gamma_{M]}\lambda_4+\frac{\tilde{t}\tilde{u}}{16}k_2^{[N}\bar{\lambda}_1\Gamma^{M]}\lambda_2 k_{3[N}\bar{\lambda}_4\Gamma_{M]}\lambda_3-\frac{\tilde{s}\tilde{u}}{16}k_2^{[M}\bar{\lambda}_2 \Gamma^{N]}\lambda_3k_{1[M}\bar{\lambda}_1\Gamma_{N]}\lambda_4\nonumber\\
&+&\frac{\tilde{t}\tilde{u}}{16}k_{1[N}\bar{\lambda}_1\Gamma_{M]}\lambda_2 k_4^{[N}\bar{\lambda}_4\Gamma^{M]}\lambda_3
-\frac{\tilde{s}\tilde{t}}{256}\bar{\lambda}_2\Gamma_M\Gamma_N(k_4 \cdot \Gamma)\lambda_3\bar{\lambda}_1(k_2 \cdot \Gamma)\Gamma^N\Gamma^M\lambda_4\nonumber\\
&+&\frac{\tilde{s}\tilde{t}}{256}\bar{\lambda_1}(k_4 \cdot \Gamma)\Gamma_M\Gamma_N\lambda_2\bar{\lambda_4}\Gamma^N\Gamma^M(k_2 \cdot \Gamma)\lambda_3-\frac{\tilde{s}\tilde{t}}{256}\bar{\lambda}_2(k_1 \cdot \Gamma)\Gamma^N\Gamma^M\lambda_3\bar{\lambda}_1\Gamma_M\Gamma_N(k_3 \cdot \Gamma)\lambda_4\nonumber\\
&+&\frac{\tilde{s}\tilde{t}}{256}\bar{\lambda_1}\Gamma^M\Gamma^N(k_3 \cdot \Gamma)\lambda_2\bar{\lambda_4}(k_1 \cdot \Gamma)\Gamma_N\Gamma_M\lambda_3-\frac{\tilde{s}\tilde{t}}{256}\bar{\lambda}_2(k_4 \cdot \Gamma)\Gamma_N\Gamma_M\lambda_3\bar{\lambda}_1(k_3 \cdot \Gamma)\Gamma^N\Gamma^M\lambda_4\nonumber\\
&+&\frac{\tilde{s}\tilde{t}}{256}\bar{\lambda_1}(k_3 \cdot \Gamma)\Gamma_N\Gamma_M\lambda_2\bar{\lambda_4}(k_2 \cdot \Gamma)\Gamma^N\Gamma^M\lambda_3-\frac{\tilde{s}\tilde{t}}{256}\bar{\lambda}_2\Gamma^M\Gamma^N(k_1 \cdot \Gamma)\lambda_3\bar{\lambda}_1\Gamma_M\Gamma_N(k_2 \cdot \Gamma)\lambda_4\nonumber\\
&+&\frac{\tilde{s}\tilde{t}}{256}\bar{\lambda_1}\Gamma^M\Gamma^N(k_4 \cdot \Gamma)\lambda_2\bar{\lambda_4}\Gamma_M\Gamma_N(k_1 \cdot \Gamma)\lambda_3 \, , \label{amplitude-full}
\end{eqnarray}
where we have used the ten-dimensional Dirac equation
\begin{eqnarray}
    (\Gamma\cdot k_i)\lambda_i=0 \, ,
\end{eqnarray}
and similarly for $\bar{\lambda_i}$. In addition, we have done some important checks on the kinematic factor of equation (\ref{amplitude-full}), namely: we have shown by explicit calculation that it satisfies the correct cyclic permutation properties and also the crossing symmetry.   

~

For certain applications, it is also convenient to consider a suitable kinematics (recall that our momenta convention corresponds to setting all incoming particles). Therefore, we set 
\begin{eqnarray}
    k_1&=&\left(\frac{\sqrt{\tilde{s}}}{2},\frac{\sqrt{\tilde{s}}}{2},0,...,0\right) \, , \nonumber \\
     k_2&=&\left(\frac{\sqrt{\tilde{s}}}{2},-\frac{\sqrt{\tilde{s}}}{2},0,...,0 \right) \, , \nonumber\\
      k_3&=&\left(-\frac{\sqrt{\tilde{s}}}{2},\frac{\sqrt{\tilde{s}}}{2}\cos\theta,\frac{\sqrt{\tilde{s}}}{2}\sin\theta,0,...,0 \right)\nonumber\\
     k_4&=&\left(-\frac{\sqrt{\tilde{s}}}{2},-\frac{\sqrt{\tilde{s}}}{2}\cos\theta,-\frac{\sqrt{\tilde{s}}}{2}\sin\theta,0,...,0 \right) \, . \label{kinematics}
\end{eqnarray}
By using this kinematics on equation (\ref{amplitude-full}), after a tedious algebraic calculation whose details are partially presented in appendix B, this equation leads to a very large expression which can be rewritten in a more compact form by regrouping certain terms. Thus, we obtain
\begin{eqnarray}
 &  & K_{\text{closed}}^{\text{4-dilatino}}(\tilde{1},\tilde{2},\tilde{3},\tilde{4} )=
\frac{\tilde{s}}{64}(7\tilde{s}\tilde{u}-\tilde{s}^2\cos\theta-2\tilde{u}^2)\bar{\lambda}_2\Gamma^0\lambda_3\bar{\lambda}_1\Gamma_{0}\lambda_4 \nonumber \\
&+&  \frac{\tilde{t}}{64}(\tilde{s}\tilde{t}\cos\theta-\tilde{s}\tilde{u}\cos\theta-8\tilde{t}\tilde{u})\bar{\lambda}_1\Gamma^0\lambda_2\bar{\lambda}_4\Gamma_{0}\lambda_3 \nonumber \\
&+& \frac{\tilde{s}}{64}(11\tilde{s}\tilde{u}-\tilde{s}^2+2\tilde{u}^2+\frac{\tilde{s}\tilde{u}}{2}\sin^2\theta)\bar{\lambda}_2\Gamma^1\lambda_3\bar{\lambda}_1\Gamma_{1}\lambda_4 \nonumber \\
&-& \frac{\tilde{t}}{64}(\tilde{s}^2\cos^2\theta+8\tilde{t}\tilde{u}) \bar{\lambda}_1\Gamma^1\lambda_2\bar{\lambda}_4\Gamma_{1}\lambda_3+ \frac{\tilde{s}}{128}(16\tilde{s}\tilde{u}-4\tilde{t}^2+7\tilde{s}\tilde{u}+\tilde{s}\tilde{u}\cos^2\theta)\bar{\lambda}_2\Gamma^2\lambda_3\bar{\lambda}_1\Gamma_{2}\lambda_4 \nonumber \\
&+& \frac{\tilde{t}}{64}(\tilde{s}^2\cos^2\theta+\tilde{s}^2-12\tilde{t}\tilde{u})\bar{\lambda}_1\Gamma^2\lambda_2\bar{\lambda}_4\Gamma_{2}\lambda_3\nonumber\\
 &+&\frac{\tilde{s}^2}{64}(12\tilde{u}+2\tilde{t})\bar{\lambda}_2\Gamma^{i_{>2}}\lambda_3\bar{\lambda}_1\Gamma_{i_{>2}}\lambda_4-\frac{\tilde{t}}{64}(-2\tilde{s}^2+12\tilde{t}\tilde{u})\bar{\lambda}_1\Gamma^{i_{>2}}\lambda_2\bar{\lambda}_4\Gamma_{i_{>2}}\lambda_3\nonumber\\
&+&\frac{\tilde{s}^2}{128}(1+\cos\theta)(\tilde{u}-\tilde{s})(\bar{\lambda}_2\Gamma^{0}\lambda_3\bar{\lambda}_1\Gamma_{1}\lambda_4+\bar{\lambda}_2\Gamma^{1}\lambda_3\bar{\lambda}_1\Gamma_{0}\lambda_4) \nonumber \\
&+&\frac{\tilde{s}^2}{128}\sin\theta(\tilde{u}-\tilde{s})(\bar{\lambda}_2\Gamma^{0}\lambda_3\bar{\lambda}_1\Gamma_{2}\lambda_4
 +\bar{\lambda}_2\Gamma^{2}\lambda_3\bar{\lambda}_1\Gamma_{0}\lambda_4)\nonumber\\
&+&\frac{\tilde{s}^2}{128}(-2\tilde{s}+\sin\theta(\tilde{s}-2\tilde{u}-\tilde{u}\cos\theta))\bar{\lambda}_2\Gamma^1\lambda_3\bar{\lambda}_1\Gamma_{2}\lambda_4-\frac{\tilde{s}^2\tilde{t}}{64}\sin\theta \cos\theta\bar{\lambda}_1\Gamma^1\lambda_2\bar{\lambda}_4\Gamma_{2}\lambda_3\nonumber\\
&-&\frac{\tilde{s}^2}{128}\sin\theta(\tilde{s}+2\tilde{u}+\tilde{u}\cos\theta)\bar{\lambda}_2\Gamma^2\lambda_3\bar{\lambda}_1\Gamma_{1}\lambda_4+\frac{\tilde{s}\tilde{t}}{64}\sin\theta(\tilde{u}-\tilde{t}(1-\cos\theta))\bar{\lambda}_1\Gamma^2\lambda_2\bar{\lambda}_4\Gamma_{1}\lambda_3\nonumber\\
&+& \frac{\tilde{s}^2}{128}\sin\theta(\tilde{s}-\tilde{u})(\bar{\lambda}_2\Gamma^0\lambda_3\bar{\lambda}_1\Gamma_{0}\Gamma_{1}\Gamma_{2}\lambda_4+\bar{\lambda}_2\Gamma^0\Gamma^1\Gamma^2\lambda_3\bar{\lambda}_1\Gamma_{0}\lambda_4)\nonumber \\
&+&\frac{\tilde{s}^2\tilde{t}}{128}\sin\theta\bar{\lambda}_2\Gamma^1\lambda_3\bar{\lambda}_1\Gamma_{0}\Gamma_{1}\Gamma_{2}\lambda_4+\frac{\tilde{s}^2}{128}\sin\theta(\tilde{s}-\tilde{u})\bar{\lambda}_2\Gamma^0\Gamma^1\Gamma^2\lambda_3\bar{\lambda}_1\Gamma_{1}\lambda_4 \nonumber \\
&+&\frac{\tilde{s}\tilde{t}}{64}\sin\theta(\tilde{u}-\tilde{t})\bar{\lambda}_1\Gamma^0\Gamma^1\Gamma^2\lambda_2\bar{\lambda}_4\Gamma_{1}\lambda_3+\frac{\tilde{s}\tilde{t}}{64}(\tilde{u}-\tilde{s})\bar{\lambda}_2\Gamma^2\lambda_3\bar{\lambda}_1\Gamma_{0}\Gamma_{1}\Gamma_{2}\lambda_4\nonumber\\
&+&\frac{\tilde{s}\tilde{t}}{64}(\tilde{u}-\tilde{t})\bar{\lambda}_1\Gamma^2\lambda_2\bar{\lambda}_4\Gamma_{0}\Gamma_{1}\Gamma_{2}\lambda_3+\frac{\tilde{s}\tilde{t}}{64}(\tilde{u}-\tilde{s})\bar{\lambda}_2\Gamma^{i_{>2}}\lambda_3\bar{\lambda}_1\Gamma_{0}\Gamma_{1}\Gamma_{i_{>2}}\lambda_4 \nonumber \\
&+&\frac{\tilde{s}\tilde{t}}{64}(\tilde{u}-\tilde{t})\bar{\lambda}_1\Gamma^{i_{>2}}\lambda_2\bar{\lambda}_4\Gamma_{0}\Gamma_{1}\Gamma_{i_{>2}}\lambda_3+\frac{\tilde{s}\tilde{t}}{64}(\tilde{u}-\tilde{s})\bar{\lambda}_2\Gamma^0\Gamma^1\Gamma^{i_{>1}}\lambda_3\bar{\lambda}_1\Gamma_{i_{>1}}\lambda_4 \nonumber \\
&+&\frac{\tilde{s}\tilde{t}}{64}\cos\theta(\tilde{t}-\tilde{u})\bar{\lambda}_1\Gamma^0\Gamma^1\Gamma^{i_{>1}}\lambda_2\bar{\lambda}_4\Gamma_{i_{>1}}\lambda_3+\frac{\tilde{s}^2}{128}\sin\theta(\tilde{u}-\tilde{s})\bar{\lambda}_2\Gamma^0\Gamma^2\Gamma^{i_{>2}}\lambda_3\bar{\lambda}_1\Gamma_{i_{>2}}\lambda_4\nonumber\\
&+&\frac{\tilde{s}\tilde{t}}{64}\sin\theta(\tilde{t}-\tilde{u})\bar{\lambda}_1\Gamma^0\Gamma^2\Gamma^{i_{>2}}\lambda_2\bar{\lambda}_4\Gamma_{i_{>2}}\lambda_3+\frac{\tilde{s}^2}{128}\sin\theta(\tilde{s}-\tilde{u})\bar{\lambda}_2\Gamma^1\Gamma^2\Gamma^{i_{>2}}\lambda_3\bar{\lambda}_1\Gamma_{i_{>2}}\lambda_4\nonumber\\
&+&\frac{\tilde{s}^2\tilde{t}}{512}\bar{\lambda}_2\Gamma^N\Gamma^P\Gamma^M\lambda_3\bar{\lambda}_1\Gamma_M\Gamma_P\Gamma_N\lambda_4-\frac{\tilde{s}\tilde{t}^2}{512}\bar{\lambda_1}\Gamma^M\Gamma^P\Gamma^N\lambda_2\bar{\lambda_4}\Gamma_N\Gamma_P\Gamma_M\lambda_3\nonumber\\
&+&\frac{\tilde{s}\tilde{t}\tilde{u}}{512}(\bar{\lambda}_2\Gamma^M\Gamma^P\Gamma^N\lambda_3\bar{\lambda}_1\Gamma_M\Gamma_P\Gamma_N\lambda_4-\bar{\lambda_1}\Gamma^M\Gamma^N\Gamma^P\lambda_2\bar{\lambda_4}\Gamma_M\Gamma_N\Gamma_P\lambda_3)\nonumber\\
&+&\frac{\tilde{s}^2\tilde{t}}{1024}[-\bar{\lambda}_2\Gamma^M\Gamma^N\Gamma^0\lambda_3\bar{\lambda}_1\Gamma_{0}\Gamma_N\Gamma_M\lambda_4+\bar{\lambda}_2\Gamma^M\Gamma^N\Gamma^0\lambda_3\bar{\lambda}_1\Gamma_{1}\Gamma_N\Gamma_M\lambda_4
\nonumber \\
&-&\cos\theta\bar{\lambda}_2\Gamma^M\Gamma^N\Gamma^1\lambda_3\bar{\lambda}_1\Gamma_{1}\Gamma_N\Gamma_M\lambda_4+\cos\theta\bar{\lambda}_2\Gamma^M\Gamma^N\Gamma^1\lambda_3\bar{\lambda}_1\Gamma_{0}\Gamma_N\Gamma_M\lambda_4 \nonumber \\
&+&\sin\theta\bar{\lambda}_2\Gamma^M\Gamma^N\Gamma^2\lambda_3\bar{\lambda}_1\Gamma_{0}\Gamma_N\Gamma_M\lambda_4-\sin\theta\bar{\lambda}_2\Gamma^M\Gamma^N\Gamma^2\lambda_3\bar{\lambda}_1\Gamma_{1}\Gamma_N\Gamma_M\lambda_4\nonumber\\
&-&\bar{\lambda}_2\Gamma^0\Gamma^N\Gamma^M\lambda_3\bar{\lambda}_1\Gamma_{M}\Gamma_N\Gamma_0\lambda_4+\cos\theta\bar{\lambda}_2\Gamma^0\Gamma^N\Gamma^M\lambda_3\bar{\lambda}_1\Gamma_{M}\Gamma_N\Gamma_1\lambda_4 \nonumber \\
&+&\sin\theta\bar{\lambda}_2\Gamma^0\Gamma^N\Gamma^M\lambda_3\bar{\lambda}_1\Gamma_{M}\Gamma_N\Gamma_2\lambda_4
+\bar{\lambda}_2\Gamma^1\Gamma^N\Gamma^M\lambda_3\bar{\lambda}_1\Gamma_{M}\Gamma_N\Gamma_0\lambda_4 \nonumber \\
&-&\cos\theta\bar{\lambda}_2\Gamma^1\Gamma^N\Gamma^M\lambda_3\bar{\lambda}_1\Gamma_{M}\Gamma_N\Gamma_1\lambda_4-\sin\theta\bar{\lambda}_2\Gamma^1\Gamma^N\Gamma^M\lambda_3\bar{\lambda}_1\Gamma_{M}\Gamma_N\Gamma_2\lambda_4\nonumber\\
&+&\bar{\lambda}_2\Gamma^0\Gamma^N\Gamma^M\lambda_3\bar{\lambda}_1\Gamma_{0}\Gamma_N\Gamma_M\lambda_4+\cos^2\theta\bar{\lambda}_2\Gamma^1\Gamma^N\Gamma^M\lambda_3\bar{\lambda}_1\Gamma_{1}\Gamma_N\Gamma_M\lambda_4 \nonumber \\
&-&\cos\theta\bar{\lambda}_2\Gamma^0\Gamma^N\Gamma^M\lambda_3\bar{\lambda}_1\Gamma_{1}\Gamma_N\Gamma_M\lambda_4
-\sin\theta\bar{\lambda}_2\Gamma^0\Gamma^N\Gamma^M\lambda_3\bar{\lambda}_1\Gamma_{2}\Gamma_N\Gamma_M\lambda_4 \nonumber \\
&-&\cos\theta\bar{\lambda}_2\Gamma^1\Gamma^N\Gamma^M\lambda_3\bar{\lambda}_1\Gamma_{0}\Gamma_N\Gamma_M\lambda_4+\sin\theta \cos\theta\bar{\lambda}_2\Gamma^1\Gamma^N\Gamma^M\lambda_3\bar{\lambda}_1\Gamma_{2}\Gamma_N\Gamma_M\lambda_4\nonumber\\
&-&\sin\theta \bar{\lambda}_2\Gamma^2\Gamma^N\Gamma^M\lambda_3\bar{\lambda}_1\Gamma_{0}\Gamma_N\Gamma_M\lambda_4+\sin\theta \cos\theta\bar{\lambda}_2\Gamma^2\Gamma^N\Gamma^M\lambda_3\bar{\lambda}_1\Gamma_{1}\Gamma_N\Gamma_M\lambda_4 \nonumber \\
&+& \sin^2\theta\bar{\lambda}_2\Gamma^2\Gamma^N\Gamma^M\lambda_3\bar{\lambda}_1\Gamma_{2}\Gamma_N\Gamma_M\lambda_4
+\bar{\lambda}_2\Gamma^M\Gamma^N\Gamma^0\lambda_3\bar{\lambda}_1\Gamma_{M}\Gamma_N\Gamma_0\lambda_4 \nonumber \\
&+&\bar{\lambda}_2\Gamma^M\Gamma^N\Gamma^1\lambda_3\bar{\lambda}_1\Gamma_{M}\Gamma_N\Gamma_1\lambda_4-\bar{\lambda}_2\Gamma^M\Gamma^N\Gamma^0\lambda_3\bar{\lambda}_1\Gamma_{M}\Gamma_N\Gamma_1\lambda_4 \nonumber \\
&-&\bar{\lambda}_2\Gamma^M\Gamma^N\Gamma^1\lambda_3\bar{\lambda}_1\Gamma_{M}\Gamma_N\Gamma_0\lambda_4 +\bar{\lambda_1}\Gamma^0\Gamma^N\Gamma^M\lambda_2\bar{\lambda_4} \Gamma_{0}\Gamma_N\Gamma_M\lambda_3 \nonumber \\
&-&\bar{\lambda_1}\Gamma^0\Gamma^N\Gamma^M\lambda_2\bar{\lambda_4} \Gamma_{1}\Gamma_N\Gamma_M\lambda_3+\cos\theta\bar{\lambda_1}\Gamma^1\Gamma^N\Gamma^M\lambda_2\bar{\lambda_4} \Gamma_{0}\Gamma_N\Gamma_M\lambda_3 \nonumber \\
&-&\cos\theta\bar{\lambda_1}\Gamma^1\Gamma^N\Gamma^M\lambda_2\bar{\lambda_4} \Gamma_{1}\Gamma_N\Gamma_M\lambda_3+\sin\theta\bar{\lambda_1}\Gamma^2\Gamma^N\Gamma^M\lambda_2\bar{\lambda_4} \Gamma_{0}\Gamma_N\Gamma_M\lambda_3\nonumber\\
&-&\sin\theta\bar{\lambda_1}\Gamma^2\Gamma^N\Gamma^M\lambda_2\bar{\lambda_4} \Gamma_{1}\Gamma_N\Gamma_M\lambda_3+\bar{\lambda_1}\Gamma_M\Gamma_N\Gamma_0\lambda_2\bar{\lambda_4} \Gamma^M\Gamma^N\Gamma^{0}\lambda_3 \nonumber \\
&+&\cos\theta\bar{\lambda_1}\Gamma_M\Gamma_N\Gamma_1\lambda_2\bar{\lambda_4} \Gamma^M\Gamma^N\Gamma^{0}\lambda_3-\sin\theta\bar{\lambda_1}\Gamma^M\Gamma^N\Gamma^2\lambda_2\bar{\lambda_4} \Gamma_M\Gamma_N\Gamma_{0}\lambda_3 \nonumber \\
&+&\bar{\lambda_1}\Gamma^M\Gamma^N\Gamma^{0}\lambda_2\bar{\lambda_4} \Gamma_M\Gamma_N\Gamma_1\lambda_3-\cos\theta\bar{\lambda_1}\Gamma^M\Gamma^N\Gamma^{1}\lambda_2\bar{\lambda_4} \Gamma_M\Gamma_N\Gamma_1\lambda_3\nonumber\\
&-&\sin\theta\bar{\lambda_1}\Gamma^M\Gamma^N\Gamma^{2}\lambda_2\bar{\lambda_4} \Gamma_M\Gamma_N\Gamma_1\lambda_3+\bar{\lambda_1}\Gamma^0\Gamma^N\Gamma^M\lambda_2\bar{\lambda_4} \Gamma_M\Gamma_N\Gamma_{0}\lambda_3 \nonumber \\
&-&\bar{\lambda_1}\Gamma^0\Gamma^N\Gamma^M\lambda_2\bar{\lambda_4} \Gamma_M\Gamma_N\Gamma_{1}\lambda_3
-\cos\theta\bar{\lambda_1}\Gamma^1\Gamma^N\Gamma^M\lambda_2\bar{\lambda_4} \Gamma_M\Gamma_N\Gamma_{0}\lambda_3\nonumber \\
&+&\cos\theta\bar{\lambda_1}\Gamma^1\Gamma^N\Gamma^M\lambda_2\bar{\lambda_4}\Gamma_M\Gamma_N \Gamma_{1}\lambda_3-\sin\theta\bar{\lambda_1}\Gamma^2\Gamma^N\Gamma^M\lambda_2\bar{\lambda_4} \Gamma_M\Gamma_N\Gamma_{0}\lambda_3\nonumber\\
&+&\sin\theta\bar{\lambda_1}\Gamma^2\Gamma^N\Gamma^M\lambda_2\bar{\lambda_4}\Gamma_M\Gamma_N \Gamma_{1}\lambda_3+\bar{\lambda_1}\Gamma^{M}\Gamma^N\Gamma^0\lambda_2\bar{\lambda_4} \Gamma_{0}\Gamma_N\Gamma_M\lambda_3 \nonumber \\
&+&\cos\theta\bar{\lambda_1}\Gamma^{M}\Gamma^N\Gamma^1\lambda_2\bar{\lambda_4} \Gamma_{0}\Gamma_N\Gamma_M\lambda_3+\sin\theta\bar{\lambda_1}\Gamma^{M}\Gamma^N\Gamma^2\lambda_2\bar{\lambda_4} \Gamma_{0}\Gamma_N\Gamma_M\lambda_3 \nonumber \\
&+&\bar{\lambda_1}\Gamma^{M}\Gamma^N\Gamma^0\lambda_2\bar{\lambda_4} \Gamma_{1}\Gamma_N\Gamma_M\lambda_3+\cos\theta\bar{\lambda_1}\Gamma^{M}\Gamma^N\Gamma^1\lambda_2\bar{\lambda_4}\Gamma_{1}\Gamma_N\Gamma_M\lambda_3\nonumber\\
  &+&\sin\theta\bar{\lambda_1}\Gamma^{M}\Gamma^N\Gamma^2\lambda_2\bar{\lambda_4} \Gamma_{1}\Gamma_N\Gamma_M\lambda_3] \, . \label{4-dilatinos-after-kinematics} \nonumber \\
  &&
\end{eqnarray}
Then, multiplying by the corresponding constants and Euler gamma functions the four-dilatino scattering amplitude in type II superstring theory becomes
\begin{eqnarray}
&&\mathcal{A}_{\text{closed}}^{\text{4-dilatinos}}(k_1,k_2,k_3,k_4) =  -4 \pi g_s^2 \alpha'^7 \ \left(\prod_{\chi=\tilde{s}, \tilde{t}, \tilde{u}} \frac{\Gamma(-\alpha' \chi/4)}{\Gamma(1+\alpha' \chi/4)}\right) \ K_{\text{closed}}^{\text{4-dilatinos}}(\tilde{1},\tilde{2},\tilde{3},\tilde{4}) \, . \nonumber \\ 
&&
\end{eqnarray}

\vspace{0.5cm}

We should emphasize that although equation (\ref{4-dilatinos-after-kinematics}) apparently shows dependence with respect to the three Mandelstam variables and also the scattering angle, it is actually a function of only two independent variables. We present this equation in this form since it is a shorter expression. By using the kinematics (\ref{kinematics}) we can write the scattering amplitude in terms of $\tilde{s}$ and the scattering angle. However, it turns out that the resulting expression contains many more terms. On the other hand, the kinematic factor (\ref{4-dilatinos-after-kinematics}) is a much shorter version, obtained at expense of reintroducing variables $\tilde{t}$ and $\tilde{u}$. To clarify this point we may present an explicit example. Let us consider the term $\mathcal{T}_{1,5}$ written only in terms of $\tilde{s}$ and the scattering angle:
\begin{eqnarray}
\mathcal{T}_{1,5}&=& -\frac{\tilde{s}^2}{32}\bar{\lambda}_2(k_4 \cdot \Gamma)\Gamma^M(k_1 \cdot
\Gamma)\lambda_3\bar{\lambda}_1\Gamma_M\lambda_4\nonumber\\ 
&=& -\frac{\tilde{s}^2}{32}[-\frac{\tilde{s}}{4}\bar{\lambda}_2\Gamma^0\lambda_3\bar{\lambda}_1\Gamma_0\lambda_4+\frac{\tilde{s}}{4}\bar{\lambda}_2 \Gamma^1\lambda_3\bar{\lambda}_1\Gamma_0\lambda_4+\frac{\tilde{s}}{4}\cos\theta\bar{\lambda}_2\Gamma^1\lambda_3\bar{\lambda}_1\Gamma_0\lambda_4\nonumber\\
&-&\frac{\tilde{s}}{4}\cos\theta\bar{\lambda}_2\Gamma^0 \lambda_3\bar{\lambda}_1\Gamma_0\lambda_4+\frac{\tilde{s}}{4}\sin\theta\bar{\lambda}_2\Gamma^2\lambda_3\bar{\lambda}_1\Gamma_0\lambda_4-\frac{\tilde{s}}{4}\sin\theta\bar{\lambda}_2\Gamma^0 \Gamma^1\Gamma^2\lambda_3\bar{\lambda}_1\Gamma_0\lambda_4\nonumber\\
&+&\frac{\tilde{s}}{4}\bar{\lambda}_2\Gamma^1 \lambda_3\bar{\lambda}_1\Gamma_1\lambda_4+\frac{\tilde{s}}{4}\bar{\lambda}_2\Gamma^2 \lambda_3\bar{\lambda}_1\Gamma_2\lambda_4+\frac{\tilde{s}}{4}\bar{\lambda}_2\Gamma^{i_{>2}} \lambda_3\bar{\lambda}_1\Gamma_{i_{>2}}\lambda_4\nonumber\\
&-&\frac{\tilde{s}}{4}\bar{\lambda}_2\Gamma^0\lambda_3\bar{\lambda}_1\Gamma_1\lambda_4-\frac{\tilde{s}}{4}\bar{\lambda}_2\Gamma^0\Gamma^1 \Gamma^2\lambda_3\bar{\lambda}_1\Gamma_2\lambda_4-\frac{\tilde{s}}{4}\bar{\lambda}_2\Gamma^0\Gamma^1\Gamma^{i_{>2}} \lambda_3\bar{\lambda}_1\Gamma_{i_{>2}}\lambda_4\nonumber\\
&-&\frac{\tilde{s}}{4}\cos\theta\bar{\lambda}_2\Gamma^0\lambda_3\bar{\lambda}_1\Gamma_1\lambda_4+\frac{\tilde{s}}{4}\cos\theta\bar{\lambda}_2\Gamma^0\Gamma^1\Gamma^2 \lambda_3\bar{\lambda}_1\Gamma_2\lambda_4+\frac{\tilde{s}}{4}\cos\theta\bar{\lambda}_2\Gamma^0\Gamma^1\Gamma^{i_{>2}} \lambda_3\bar{\lambda}_1\Gamma_{i_{>2}}\lambda_4\nonumber\\
&+&\frac{\tilde{s}}{4}\cos\theta\bar{\lambda}_2 \Gamma^1\lambda_3\bar{\lambda}_1\Gamma_1\lambda_4-\frac{\tilde{s}}{4}\cos\theta\bar{\lambda}_2\Gamma^2 \lambda_3\bar{\lambda}_1\Gamma_2\lambda_4-\frac{\tilde{s}}{4}\cos\theta\bar{\lambda}_2\Gamma^{i_{>2}} \lambda_3\bar{\lambda}_1\Gamma_{i_{>2}}\lambda_4\nonumber\\
&-&\frac{\tilde{s}}{4}\sin\theta\bar{\lambda}_2\Gamma^0\Gamma^1 \Gamma^2\lambda_3\bar{\lambda}_1\Gamma_1\lambda_4-\frac{\tilde{s}}{4}\sin\theta\bar{\lambda}_2 \Gamma^0\lambda_3\bar{\lambda}_1\Gamma_2\lambda_4+\frac{\tilde{s}}{4}\sin\theta\bar{\lambda}_2\Gamma^0\Gamma^2\Gamma^{i_{>2}}  \lambda_3\bar{\lambda}_1\Gamma_{i_{>2}} \lambda_4\nonumber\\ 
&+&\frac{\tilde{s}}{4}\sin\theta\bar{\lambda}_2\Gamma^2\lambda_3\bar{\lambda}_1\Gamma_1\lambda_4+\frac{\tilde{s}}{4}\sin\theta\bar{\lambda}_2 \Gamma^1\lambda_3\bar{\lambda}_1\Gamma_2\lambda_4-\frac{\tilde{s}}{4}\sin\theta\bar{\lambda}_2\Gamma^1\Gamma^2\Gamma^{i_{>2}} \lambda_3\bar{\lambda}_1\Gamma_{i_{>2}}\lambda_4] \, , \nonumber \\
&&
\end{eqnarray}
with 24 terms, which can be compared with the expression that we obtain in terms of the three Mandelstam variables and the scattering angle:
\begin{eqnarray}
\mathcal{T}_{1,5}&=& -\frac{\tilde{s}^2}{32}[\frac{\tilde{u}}{2}(\bar{\lambda}_2\Gamma^0\lambda_3\bar{\lambda}_1\Gamma_0\lambda_4-\bar{\lambda}_2 \Gamma^1\lambda_3\bar{\lambda}_1\Gamma_0\lambda_4-\bar{\lambda}_2\Gamma^1 \lambda_3\bar{\lambda}_1\Gamma_1\lambda_4+\bar{\lambda}_2\Gamma^0\lambda_3\bar{\lambda}_1\Gamma_1\lambda_4)
\nonumber \\
&+&\frac{\tilde{t}}{2}(-\bar{\lambda}_2\Gamma^2 \lambda_3\bar{\lambda}_1\Gamma_2\lambda_4-\bar{\lambda}_2\Gamma^{i_{>2}} \lambda_3\bar{\lambda}_1\Gamma_{i_{>2}}\lambda_4+\bar{\lambda}_2\Gamma^0\Gamma^1 \Gamma^2\lambda_3\bar{\lambda}_1\Gamma_2\lambda_4 \nonumber \\
&+&\bar{\lambda}_2\Gamma^0\Gamma^1\Gamma^{i_{>2}} \lambda_3\bar{\lambda}_1\Gamma_{i_{>2}}\lambda_4) +\frac{\tilde{s}}{4}\sin\theta (\bar{\lambda}_2\Gamma^2\lambda_3\bar{\lambda}_1\Gamma_0\lambda_4 
- \bar{\lambda}_2\Gamma^0\Gamma^1 \Gamma^2\lambda_3\bar{\lambda}_1\Gamma_0\lambda_4 \nonumber \\
&-&\bar{\lambda}_2\Gamma^0\Gamma^1 \Gamma^2\lambda_3\bar{\lambda}_1\Gamma_1\lambda_4-\bar{\lambda}_2 \Gamma^0\lambda_3\bar{\lambda}_1\Gamma_2\lambda_4+\bar{\lambda}_2\Gamma^0\Gamma^2\Gamma^{i_{>2}}  \lambda_3\bar{\lambda}_1\Gamma_{i_{>2}} \lambda_4\nonumber\\ 
&+&\bar{\lambda}_2\Gamma^2\lambda_3\bar{\lambda}_1\Gamma_1\lambda_4+\bar{\lambda}_2 \Gamma^1\lambda_3\bar{\lambda}_1\Gamma_2\lambda_4-\bar{\lambda}_2\Gamma^1\Gamma^2\Gamma^{i_{>2}} \lambda_3\bar{\lambda}_1\Gamma_{i_{>2}}\lambda_4)] \, , 
\end{eqnarray}
which has 16 terms. By adding all the terms of the tensor product $\mathcal{T}_{i,j}$ this effect largely amplifies. A further possible simplification of these expressions may occur using the Fierz identities, though we have not used them.

%
%
\section{The high energy behavior}
%
%

Let us consider the Regge limit, i.e. $\tilde{s} \gg |\tilde{t}|$. The relevant terms contributing to the Regge behavior of the kinematic factor of the four-dilatino scattering amplitude are only the following ones:
\begin{eqnarray}
     {\cal {T}}_{1,2}^{\text{Regge}}&=&\frac{1}{8}\tilde{s}^2\tilde{u}\bar{\lambda_2}\Gamma^M\lambda_3\bar{\lambda_1}\Gamma_M\lambda_4 \, ,
\end{eqnarray}
\begin{eqnarray}
{\cal {T}}_{1,4}^{\text{Regge}}=\frac{\tilde{s}^3}{64}[\bar{\lambda}_2\Gamma^0\lambda_3\bar{\lambda}_1 \Gamma_0\lambda_4-\bar{\lambda}_2\Gamma^1\lambda_3\bar{\lambda}_1\Gamma_1\lambda_4+\bar{\lambda}_2\Gamma^1\lambda_3\bar{\lambda}_1\Gamma_0\lambda_4-\bar{\lambda}_2\Gamma^0\lambda_3\bar{\lambda}_1 \Gamma_1\lambda_4] \, ,
\end{eqnarray}
\begin{eqnarray}
{\cal {T}}_{1,5}^{\text{Regge}}= \frac{\tilde{s}^3}{64}[\bar{\lambda}_2\Gamma^0\lambda_3\bar{\lambda}_1\Gamma_0\lambda_4-\bar{\lambda}_2 \Gamma^1\lambda_3\bar{\lambda}_1\Gamma_0\lambda_4-\bar{\lambda}_2\Gamma^1 \lambda_3\bar{\lambda}_1\Gamma_1\lambda_4+\bar{\lambda}_2\Gamma^0\lambda_3\bar{\lambda}_1\Gamma_1\lambda_4] \, ,
\end{eqnarray}
\begin{eqnarray}
{\cal {T}}_{1,6}^{\text{Regge}}&=& -\frac{\tilde{s}^3}{128}[\bar{\lambda}_2\Gamma^1\lambda_3\bar{\lambda}_1\Gamma_1\lambda_4+\bar{\lambda}_2\Gamma^1\lambda_3\bar{\lambda}_1\Gamma_0\lambda_4+\bar{\lambda}_2\Gamma^0\lambda_3\bar{\lambda}_1\Gamma_1\lambda_4\nonumber\\
&+&\bar{\lambda}_2\Gamma^0\lambda_3\bar{\lambda}_1\Gamma_0\lambda_4+2\bar{\lambda}_2\Gamma^{i_{>1}}\lambda_3\bar{\lambda}_1\Gamma_{i_{>1}}\lambda_4] \, ,
\end{eqnarray}
\begin{eqnarray}
{\cal {T}}_{1,7}^{\text{Regge}}&=&-\frac{\tilde{s}^3}{128}[\bar{\lambda}_2\Gamma^1\lambda_3\bar{\lambda}_1\Gamma_1\lambda_4+\bar{\lambda}_2\Gamma^1\lambda_3\bar{\lambda}_1\Gamma_0\lambda_4+\bar{\lambda}_2\Gamma^0\lambda_3\bar{\lambda}_1\Gamma_1\lambda_4+\bar{\lambda}_2\Gamma^0\lambda_3\bar{\lambda}_1\Gamma_0\lambda_4 \nonumber \\%
&+&2\bar{\lambda}_2\Gamma^{i_{>1}}\lambda_3 \bar{\lambda}_1\Gamma_{i_{>1}}\lambda_4] \nonumber \\
\end{eqnarray}
\begin{eqnarray}
{\cal {T}}_{1,8}^{\text{Regge}}&=& -\frac{\tilde{s}^3}{64}[\bar{\lambda}_2\Gamma^0\lambda_3\bar{\lambda}_1\Gamma_0\lambda_4-\bar{\lambda}_2 \Gamma^1\lambda_3\bar{\lambda}_1\Gamma_0\lambda_4-\bar{\lambda}_2\Gamma^1 \lambda_3\bar{\lambda}_1\Gamma_1\lambda_4
+\bar{\lambda}_2\Gamma^0\lambda_3\bar{\lambda}_1\Gamma_1\lambda_4] \nonumber \\
\end{eqnarray}
\begin{eqnarray}
{\cal {T}}_{1,9}^{\text{Regge}}= \frac{\tilde{s}^3}{64}[-\bar{\lambda}_2\Gamma^0\lambda_3\bar{\lambda}_1 \Gamma_0\lambda_4+\bar{\lambda}_2\Gamma^0\lambda_3\bar{\lambda}_1 \Gamma_1\lambda_4+\bar{\lambda}_2\Gamma^1\lambda_3\bar{\lambda}_1\Gamma_1\lambda_4
     -\bar{\lambda}_2\Gamma^1\lambda_3\bar{\lambda}_1\Gamma_0\lambda_4] \, ,
\end{eqnarray}
\begin{eqnarray}
{\cal {T}}_{1,10}^{\text{Regge}}&=& -\frac{\tilde{s}^3}{128}[\bar{\lambda}_2\Gamma^{1}\lambda_3 \bar{\lambda}_1\Gamma_{1}\lambda_4+\bar{\lambda}_2\Gamma^{1}\lambda_3 \bar{\lambda}_1\Gamma_{0}\lambda_4+\bar{\lambda}_2\Gamma^{0}\lambda_3 \bar{\lambda}_1\Gamma_{1}\lambda_4\nonumber\\
&+&\bar{\lambda}_2\Gamma^{0}\lambda_3 \bar{\lambda}_1\Gamma_{0}\lambda_4+2\bar{\lambda}_2\Gamma^{i_{>1}}\lambda_3\bar{\lambda}_1\Gamma_{i_{>1}}\lambda_4] \, ,
\end{eqnarray}
\begin{eqnarray}
{\cal {T}}_{1,11}^{\text{Regge}}&=& -\frac{\tilde{s}^3}{128}[\bar{\lambda}_2\Gamma^{1}\lambda_3 \bar{\lambda}_1\Gamma_{1}\lambda_4+\bar{\lambda}_2\Gamma^{1}\lambda_3 \bar{\lambda}_1\Gamma_{0}\lambda_4+\bar{\lambda}_2\Gamma^{0}\lambda_3 \bar{\lambda}_1\Gamma_{1}\lambda_4\nonumber\\
&+&\bar{\lambda}_2\Gamma^{0}\lambda_3 \bar{\lambda}_1\Gamma_{0}\lambda_4
+2\bar{\lambda}_2\Gamma^{i_{>1}}\lambda_3\bar{\lambda}_1\Gamma_{i_{>1}}\lambda_4] \, .
\end{eqnarray}
We should notice that terms with $\Gamma^9$ vanish since we consider a ten-dimensional right-handed dilatino. Therefore, adding these relevant terms in the Regge limit we obtain:
\begin{eqnarray}
&& K_{\text{closed}}^{\text{4-dilatino-Regge}}(\tilde{1},\tilde{2},\tilde{3},\tilde{4})=-\frac{\tilde{s}^3}{32}[5(\bar{\lambda}_2\Gamma^{0}\lambda_3 \bar{\lambda}_1\Gamma_{0}\lambda_4+\bar{\lambda}_2\Gamma^{1}\lambda_3 \bar{\lambda}_1\Gamma_{1}\lambda_4) \nonumber \\
&+&\bar{\lambda}_2\Gamma^{0}\lambda_3\bar{\lambda}_1\Gamma_{1}\lambda_4+\bar{\lambda}_2\Gamma^{1}\lambda_3\bar{\lambda}_1\Gamma_{0}\lambda_4+6\bar{\lambda}_2\Gamma^{i_{>1}}\lambda_3\bar{\lambda}_1\Gamma_{{i_{>1}}}\lambda_4] \, .
\end{eqnarray}
The corresponding Regge limit of the four-dilatino scattering amplitude is given by 
\begin{eqnarray}
&&\mathcal{A}_{\text{closed}}^{\text{4-dilatinos-Regge}}(k_1,k_2,k_3,k_4) = -4\pi g_s^2 \alpha'^7 \ K_{\text{closed}}^{\text{4-dilatinos-Regge}}(1,2,3,4) \times \nonumber \\
&& \,\,\,\,\,\,\,\,\,\,\,\,\,\,\,\,\,\,\,\,\,\, \frac{\sin[\frac{\pi\alpha'}{4}(\tilde{s}+\tilde{t})]}{\sin(\frac{\pi\alpha'\tilde{s}}{4})} \left(\frac{\alpha'\tilde{s}}{4}\right)^{-2+\frac{\alpha'\tilde{t}}{2}} \ e^{2-\frac{\alpha'\tilde{t}}{2}}\frac{\Gamma(-\alpha'\tilde{t}/4)}{\Gamma(1+\alpha'\tilde{t}/4)} \nonumber \\ 
&=& \frac{\pi g_s^2 \alpha'^4}{8} (\alpha'\tilde{s})^3[5(\bar{\lambda}_2\Gamma^{0}\lambda_3 \bar{\lambda}_1\Gamma_{0}\lambda_4+\bar{\lambda}_2\Gamma^{1}\lambda_3 \bar{\lambda}_1\Gamma_{1}\lambda_4) \nonumber \\
&+&\bar{\lambda}_2\Gamma^{0}\lambda_3\bar{\lambda}_1\Gamma_{1}\lambda_4+\bar{\lambda}_2\Gamma^{1}\lambda_3\bar{\lambda}_1\Gamma_{0}\lambda_4+6\bar{\lambda}_2\Gamma^{i_{>1}}\lambda_3\bar{\lambda}_1\Gamma_{{i_{>1}}}\lambda_4]\frac{\sin[\frac{\pi\alpha'}{4}(\tilde{s}+\tilde{t})]}{\sin(\frac{\pi\alpha'\tilde{s}}{4})} \left(\frac{\alpha'\tilde{s}}{4}\right)^{-2+\frac{\alpha'\tilde{t}}{2}}\nonumber\\
&\times& e^{2-\frac{\alpha'\tilde{t}}{2}}\frac{\Gamma(-\alpha'\tilde{t}/4)}{\Gamma(1+\alpha'\tilde{t}/4)} \, ,
\end{eqnarray}
leading to
\begin{eqnarray}
&&\mathcal{A}_{\text{closed}}^{\text{4-dilatinos-Regge}}(k_1,k_2,k_3,k_4) = 2^{1-\alpha'\tilde{t}} \pi g_s^2 \alpha'^4 (\alpha'\tilde{s})^{1+\frac{\alpha'\tilde{t}}{2}}[5(\bar{\lambda}_2\Gamma^{0}\lambda_3 \bar{\lambda}_1\Gamma_{0}\lambda_4+\bar{\lambda}_2\Gamma^{1}\lambda_3 \bar{\lambda}_1\Gamma_{1}\lambda_4) \nonumber \\
&+&\bar{\lambda}_2\Gamma^{0}\lambda_3\bar{\lambda}_1\Gamma_{1}\lambda_4+\bar{\lambda}_2\Gamma^{1}\lambda_3\bar{\lambda}_1\Gamma_{0}\lambda_4+6\bar{\lambda}_2\Gamma^{i_{>1}}\lambda_3\bar{\lambda}_1\Gamma_{{i_{>1}}}\lambda_4]\frac{\sin[\frac{\pi\alpha'}{4}(\tilde{s}+\tilde{t})]}{\sin(\frac{\pi\alpha'\tilde{s}}{4})} e^{2-\frac{\alpha'\tilde{t}}{2}}\frac{\Gamma(-\alpha'\tilde{t}/4)}{\Gamma(1+\alpha'\tilde{t}/4)} \, . \label{Regge-four-dilatinos} \nonumber \\ 
\end{eqnarray}
%

%
%
\section{Conclusions}
%
%

We have explicitly obtained the four-dilatino scattering amplitude in type II superstring theory by using the Kawai-Lewellen-Tye relations. We have explicitly checked that this four-dilatino scattering amplitude is invariant under crossing symmetry, as well as the $S_4$ cyclic permutation symmetry, as expected.

In addition, we have derived the corresponding Regge behavior in the high-energy limit $\tilde{s} \gg |\tilde{t}|$ of the four-dilatino scattering amplitude in type II superstring theory. We should emphasize that the scattering amplitude $\mathcal{A}_{\text{closed}}^{\text{4-dilatinos-Regge}}(k_1,k_2,k_3,k_4)$ of equation (\ref{Regge-four-dilatinos}) is an exact result in the Regge limit. It is interesting to compare  $\mathcal{A}_{\text{closed}}^{\text{4-dilatinos-Regge}}(k_1,k_2,k_3,k_4)$ with the corresponding one of the Regge limit of four dilatons (\ref{Four-dilaton-Regge}). While in the case of four dilatons the Regge limit gives the factor $(\alpha'\tilde{s})^{2+\frac{\alpha'\tilde{t}}{2}}$, in the case of four dilatinos the factor becomes $(\alpha'\tilde{s})^{1+\frac{\alpha'\tilde{t}}{2}}$. In order to understand this difference in both scattering amplitudes, let us consider the full Euler gamma-function factor of these scattering amplitudes:
\begin{eqnarray}
&& G(\tilde{s}, \tilde{t}, \tilde{u}) =  \left(\prod_{\chi=\tilde{s}, \tilde{t}, \tilde{u}} \frac{\Gamma(-\alpha' \chi/4)}{\Gamma(1+\alpha' \chi/4)}\right) \, , \nonumber
\end{eqnarray}
that, after using the Stirling's formula for the high-energy limit $\tilde{s} \gg |\tilde{t}|$, scales as $\tilde{s}^{-2}$. The scattering amplitudes contain this Euler gamma-function factor multiplied by a kinematic factor corresponding to closed strings, which in turn is given by the product of two open-string kinematic factors. At this point we have to analyze separately both situations. In the case of four dilatons the kinematic factor behaves like $K_{\text{closed}}^{\text{4-dilatons-Regge}}(1,2,3,4) \propto \tilde{s}^4$. On the other hand, in the case of four dilatinos the kinematic factor is $K_{\text{closed}}^{\text{4-dilatino-Regge}}(\tilde{1},\tilde{2},\tilde{3},\tilde{4}) \propto \tilde{s}^3$. Notice that in the case of four fermions this is a naive power counting as  explained below. Thus, we understand the power two of the center-of-mass energy in the case of four-dilaton scattering amplitude in the Regge limit, and the power one of $\tilde{s}$ in the case of four dilatinos.
However, the power counting of $\mathcal{A}_{\text{closed}}^{\text{4-dilatinos-Regge}}(k_1,k_2,k_3,k_4)$ is more subtle. In fact, there is an additional factor $\tilde{s}$, which comes from the factor
\begin{eqnarray}
[5(\bar{\lambda}_2\Gamma^{0}\lambda_3 \bar{\lambda}_1\Gamma_{0}\lambda_4+\bar{\lambda}_2\Gamma^{1}\lambda_3 \bar{\lambda}_1\Gamma_{1}\lambda_4) + \bar{\lambda}_2\Gamma^{0}\lambda_3\bar{\lambda}_1\Gamma_{1}\lambda_4+\bar{\lambda}_2\Gamma^{1}\lambda_3\bar{\lambda}_1\Gamma_{0}\lambda_4+6\bar{\lambda}_2\Gamma^{i_{>1}}\lambda_3\bar{\lambda}_1\Gamma_{{i_{>1}}}\lambda_4] \, , \nonumber 
\label{dimensions_of_lambda}
\end{eqnarray}
in equation (\ref{Regge-four-dilatinos}). Let us explain how it occurs focusing on the dilatino wave functions. Firstly, let us recall that in the Green-Schwarz formalism the vertices for massless particles corresponding to open string states have been derived in the light-cone gauge. In particular, the vertex which accounts for the emission of a massless fermion with a wave function $u^{\dot{a}}$ and momentum $k^\mu$ is given by \cite{Green:1987sp}
\begin{equation}
V_F(u, k) = u \ F \ e^{i k \cdot X} = (u^a \ F^a + u^{\dot{a}} \ F^{\dot{a}}) \ e^{i k \cdot X} \, ,
\end{equation}
with
\begin{equation}
F^a =  (p^+/2)^{1/2} S^a \, ,
\end{equation}
while the prefactor of $F^{\dot{a}}$ is $(2 p^+)^{-1/2}$, where $p^+ \propto {\tilde{s}}^{1/2}$ (with the kinematics defined in section 3). In addition, $S^a$'s are dimensionless since they satisfy
\begin{eqnarray}
\{S_m^a, S_n^b\} = \delta_{m+n} \delta^{ab} \, .
\end{eqnarray}
Then, for each fermionic open string emission vertex there is a factor  $\tilde{s}^{1/4}$ coming from $F^a$, which in the case of four fermions leads to the factor $\tilde{s}$.  In addition, each dilatino wave function has a $\tilde{s}^{1/4}$ prefactor, which in the case of four dilatinos gives another $\tilde{s}$. These factors precisely come from the terms in (\ref{dimensions_of_lambda}). Recall that using the KLT construction, the bosonic kinematic factor carries $\tilde{s}^2$. Multiplying all these factors, it leads to $(\alpha'\tilde{s})^{2+\frac{\alpha'\tilde{t}}{2}}$ in the Regge limit.

The conclusion is that after taking into account all these elements both scattering amplitudes  $\mathcal{A}_{\text{closed}}^{\text{4-dilatons-Regge}}(k_1,k_2,k_3,k_4)$ and $\mathcal{A}_{\text{closed}}^{\text{4-dilatinos-Regge}}(k_1,k_2,k_3,k_4)$ have the same scaling behavior $(\alpha'\tilde{s})^{2+\frac{\alpha'\tilde{t}}{2}}$ in the Regge limit. Thus, this trajectory  $J(\tilde{t})=2+\frac{\alpha'\tilde{t}}{2}$ shows the Reggeization of the graviton in both cases.

~

We hope that the four spin-1/2 fermions scattering amplitudes which we have thoroughly derived from type II superstring theories for closed strings will be useful for interesting applications. For instance, it has been recently used in the calculation of the elastic four spin-1/2 fermions scattering cross section in a holographic dual confining gauge field theory \cite{Martin:2024jpe}, using the methods developed by Polchinski and Strassler \cite{Polchinski:2001tt}. Polchinski and Strassler further developed these methods to study deep inelastic scattering (DIS) \cite{Polchinski:2002jw} testing glueball. Also, mesons \cite{Koile:2014vca} and spin-1/2 fermions \cite{Kovensky:2018xxa} were investigated. Furthermore, the development of the Brower-Polchinski-Strassler-Tan pomeron \cite{Brower:2006ea} allowed to describe the $F_2$ structure function of fermions in DIS and compared it very successfully with experimental data \cite{Brower:2010wf,Jorrin:2022lua,Borsa:2023tqr}. Moreover, the development of the holographic A$_4$ pomeron in \cite{Kovensky:2018xxa} gives an excellent comparison with existing experimental data for polarized DIS of the proton helicity structure function $g_1$ as shown in \cite{Kovensky:2018xxa,Jorrin:2022lua,Borsa:2023tqr}, as well as it provides sharp predictions of  $g_1$ for the forthcoming Electron-Ion Collider.

\vspace{1.cm}

%
\centerline{\large{\bf Acknowledgments}}
%

~

 We thank Adri\'an Lugo, Carmen A. N\'u\~nez and Carlos N\'u\~nez for conversations and useful comments. We also especially thank Carmen N\'u\~nez for calling our attention on some important references during the early stage of this project. This work has been supported in part by the Consejo Nacional de Investigaciones Cient\'{\i}fi\-cas y T\'ecnicas of Argentina (CONICET).

\pagebreak

\newpage

\begin{appendices}

\section{The 30 tensor-product terms contributing to four-dilatino scattering amplitude}

We list the set of 30 tensor-product terms which appear from the kinematic factor of the closed string scattering amplitude $ K_{\text{open}}^{\text{fermionic}}(\tilde{1},\tilde{2}, \tilde{3},\tilde{4})\otimes \frac{1}{16} K_{\text{open}}^{\text{bosonic}}(1,2,4,3)$.
We firstly list the terms of the form ${\cal {T}}_{1,i}$.
\begin{eqnarray}
    {\cal {T}}_{1,1}&=&\left(-\frac{1}{8} \tilde{s} \bar{u}_2 \Gamma^{M'} u_{3} \bar{u}_1\Gamma_{M'} u_4\right) \otimes \left(-\frac{1}{16\cdot 4}\tilde{s}\tilde{t}\zeta_1^M \zeta_{3,M} \zeta_4^Q \zeta_{2,Q}\right)\nonumber\\
    &=& \frac{\tilde{s}^2\tilde{t}}{512}\bar{\lambda}_2\Gamma_N\Gamma_P\Gamma_M\lambda_3\bar{\lambda}_1\Gamma^M\Gamma^P\Gamma^M\lambda_4 \, .
\end{eqnarray}
\begin{eqnarray}
     {\cal {T}}_{1,2}&=&\left(-\frac{1}{8} \tilde{s} \bar{u}_2 \Gamma^{M'} u_{3} \bar{u}_1\Gamma_{M'} u_4\right)\otimes  \left(-\frac{1}{16\cdot 4}\tilde{s}\tilde{u}\zeta_1^M \zeta_{4,M} \zeta_3^P \zeta_{2,P}  \right) \nonumber\\
    &=&\frac{1}{8}\tilde{s}^2\tilde{u}\bar{\lambda_2}\Gamma^M\lambda_3\bar{\lambda_1}\Gamma_M\lambda_4 \, .
\end{eqnarray}
\begin{eqnarray}
{\cal {T}}_{1,3}&=&\left(-\frac{1}{8} \tilde{s} \bar{u}_2 \Gamma^{M'} u_{3} \bar{u}_1  \Gamma_{M'} u_4\right)\otimes\left(-\frac{1}{16\cdot 4}\tilde{t}\tilde{u}\zeta_1^M \zeta_{2,M} \zeta_3^P \zeta_{4,P}\right) \nonumber\\
&=& \frac{\tilde{s}\tilde{t}\tilde{u}}{512}\bar{\lambda}_2\Gamma_M\Gamma_P\Gamma_N\lambda_3\bar{\lambda}_1\Gamma^M\Gamma^P\Gamma^N\lambda_4 \, ,
\end{eqnarray}
\begin{eqnarray}
{\cal {T}}_{1,4}&=&\left(-\frac{1}{8} \tilde{s} \bar{u}_2 \Gamma^{M'} u_{3} \bar{u}_1  \Gamma_{M'} u_4\right)\otimes\left(\frac{1}{16\cdot 2}\tilde{s}\zeta_1^M k_{3,M} \zeta_4^Qk_{2,Q}\zeta_3^P\zeta_{2,P}\right)\nonumber\\
     &=& -\frac{\tilde{s}^2}{32}\bar{\lambda}_2\Gamma_M\lambda_3\bar{\lambda}_1(k_3\cdot \Gamma)\Gamma^M(k_2\cdot\Gamma)\lambda_4 \, ,
\end{eqnarray}
\begin{eqnarray}
{\cal {T}}_{1,5}&=&\left(-\frac{1}{8} \tilde{s} \bar{u}_2 \Gamma^{M'} u_{3} \bar{u}_1  \Gamma_{M'} u_4\right)\otimes\left(\frac{1}{16\cdot 2}\tilde{s}\zeta_2^N k_{4,N} \zeta_3^P k_{1,P}\zeta_1^M\zeta_{4,M}\right)\nonumber\\
     &=& -\frac{\tilde{s}^2}{32}\bar{\lambda}_2(k_4 \cdot \Gamma)\Gamma^M(k_1 \cdot \Gamma)\lambda_3\bar{\lambda}_1\Gamma_M\lambda_4 \, ,
\end{eqnarray}
\begin{eqnarray}
{\cal {T}}_{1,6}&=&\left(-\frac{1}{8} \tilde{s} \bar{u}_2 \Gamma^{M'} u_{3} \bar{u}_1  \Gamma_{M'} u_4\right)\otimes\left(\frac{1}{16\cdot 2}\tilde{s}\zeta_1^M k_{4,M} \zeta_3^P k_{2,P}\zeta_4^Q\zeta_{2,Q}\right)\nonumber\\
   &=& -\frac{\tilde{s}^2}{256}\bar{\lambda}_2\Gamma^N\Gamma^M(k_2\cdot \Gamma)\lambda_3\bar{\lambda}_1(k_4\cdot \Gamma)\Gamma_M\Gamma_N\lambda_4\nonumber\\
   &=& -\frac{\tilde{s}^2}{16}k_2^{[M}\bar{\lambda}_2\Gamma^{N]}\lambda_3 k_{4[M}\bar{\lambda}_1\Gamma_{N]}\lambda_4 \, ,
\end{eqnarray}
\begin{eqnarray}
{\cal {T}}_{1,7}&=&\left(-\frac{1}{8} \tilde{s} \bar{u}_2 \Gamma^{M'} u_{3} \bar{u}_1  \Gamma_{M'} u_4\right)\otimes\left(\frac{1}{16\cdot 2}\tilde{s}\zeta_2^N k_{3,N} \zeta_4^Q k_{1,Q}\zeta_1^M\zeta_{3,M}\right)\nonumber\\
   &=& -\frac{\tilde{s}^2}{256}\bar{\lambda}_2(k_3\cdot \Gamma)\Gamma_N\Gamma_M\lambda_3\bar{\lambda}_1\Gamma^M\Gamma^N(k_1\cdot \Gamma)\lambda_4 \nonumber \\
&=& -\frac{\tilde{s}^2}{16} k_{3[M} \bar{\lambda}_2 \Gamma_{N]} \lambda_3 k_1^{[M}\bar{\lambda}_1\Gamma^{N]}\lambda_4 \, ,
\end{eqnarray}
\begin{eqnarray}
{\cal {T}}_{1,8}&=&\left(-\frac{1}{8} \tilde{s} \bar{u}_2 \Gamma^{M'} u_{3} \bar{u}_1  \Gamma_{M'} u_4\right)\otimes\left(\frac{1}{16\cdot 2}\tilde{u}\zeta_2^N k_{1,N} \zeta_3^P k_{4,P}\zeta_4^Q\zeta_{1,Q}\right)\nonumber\\
    &=&-\frac{\tilde{s}\tilde{u}}{32}\bar{\lambda}_2(k_1 \cdot \Gamma)\Gamma^M(k_4 \cdot \Gamma)\lambda_3\bar{\lambda}_1\Gamma_M\lambda_4 \, ,
\end{eqnarray}
\begin{eqnarray}
{\cal {T}}_{1,9}&=&\left(-\frac{1}{8} \tilde{s} \bar{u}_2 \Gamma^{M'} u_{3} \bar{u}_1  \Gamma_{M'} u_4\right)\otimes\left(\frac{1}{16\cdot 2}\tilde{u}\zeta_4^N k_{3,N} \zeta_1^M k_{2,M}\zeta_2^Q\zeta_{3,Q}\right)\nonumber\\
    &=&-\frac{\tilde{s}\tilde{u}}{32}\bar{\lambda}_2\Gamma_M\lambda_3\bar{\lambda}_1(k_2 \cdot \Gamma)\Gamma^M(k_3 \cdot \Gamma)\lambda_4 \, ,
\end{eqnarray}
\begin{eqnarray}
{\cal {T}}_{1,10}&=&\left(-\frac{1}{8} \tilde{s} \bar{u}_2 \Gamma^{M'} u_{3} \bar{u}_1  \Gamma_{M'} u_4\right)\otimes\left(\frac{1}{16\cdot 2}\tilde{u}\zeta_2^N k_{3,N} \zeta_1^M k_{4,M}\zeta_4^Q\zeta_{3,Q}\right)\nonumber\\
    &=& -\frac{\tilde{s}\tilde{u}}{256}\bar{\lambda}_2(k_3 \cdot \Gamma)\Gamma^M\Gamma^N\lambda_3\bar{\lambda}_1(k_4\cdot \Gamma)\Gamma_M\Gamma_N\lambda_4\nonumber\\
    &=& -\frac{\tilde{s}\tilde{u}}{16}k_3^{[N}\bar{\lambda}_2 \Gamma^{M]}\lambda_3k_{4[N}\bar{\lambda}_1\Gamma_{M]}\lambda_4 \, ,
\end{eqnarray}
\begin{eqnarray}
{\cal {T}}_{1,11}&=&\left(-\frac{1}{8} \tilde{s} \bar{u}_2 \Gamma^{M'} u_{3} \bar{u}_1  \Gamma_{M'} u_4\right)\otimes\left(\frac{1}{16\cdot 2}\tilde{u}\zeta_4^Q k_{1,Q} \zeta_3^P k_{2,P}\zeta_2^N\zeta_{1,N}\right)\nonumber\\
&=& -\frac{\tilde{s}\tilde{u}}{256}\bar{\lambda}_2\Gamma^M\Gamma^N(k_2 \cdot \Gamma)\lambda_3\bar{\lambda}_1\Gamma_M\Gamma_N(k_1\cdot \Gamma)\lambda_4\nonumber\\
&=& -\frac{\tilde{s}\tilde{u}}{16}k_2^{[M}\bar{\lambda}_2 \Gamma^{N]}\lambda_3k_{1[M}\bar{\lambda}_1\Gamma_{N]}\lambda_4 \, ,
\end{eqnarray}
\begin{eqnarray}
{\cal {T}}_{1,12}&=&\left(-\frac{1}{8} \tilde{s} \bar{u}_2 \Gamma^{M'} u_{3} \bar{u}_1  \Gamma_{M'} u_4\right)\otimes\left(\frac{1}{16\cdot 2}\tilde{t}\zeta_1^M k_{2,M} \zeta_3^P k_{4,P}\zeta_4^Q\zeta_{2,Q}\right)\nonumber\\
   &=& -\frac{\tilde{s}\tilde{t}}{256}\bar{\lambda}_2\Gamma_M\Gamma_N(k_4 \cdot \Gamma)\lambda_3\bar{\lambda}_1(k_2 \cdot \Gamma)\Gamma^N\Gamma^M\lambda_4 \, ,
\end{eqnarray}
\begin{eqnarray}
{\cal {T}}_{1,13}&=&\left(-\frac{1}{8} \tilde{s} \bar{u}_2 \Gamma^{M'} u_{3} \bar{u}_1  \Gamma_{M'} u_4\right)\otimes\left(\frac{1}{16\cdot 2}\tilde{t}\zeta_4^Q k_{3,Q} \zeta_2^N k_{1,N}\zeta_1^M\zeta_{3,M}\right)\nonumber\\
  &=& -\frac{\tilde{s}\tilde{t}}{256}\bar{\lambda}_2(k_1 \cdot \Gamma)\Gamma^N\Gamma^M\lambda_3\bar{\lambda}_1\Gamma_M\Gamma_N(k_3 \cdot \Gamma)\lambda_4 \, ,
\end{eqnarray}
\begin{eqnarray}
{\cal {T}}_{1,14}&=& \left(-\frac{1}{8} \tilde{s} \bar{u}_2 \Gamma^{M'} u_{3} \bar{u}_1  \Gamma_{M'} u_4\right)\otimes\left(\frac{1}{16\cdot 2}\tilde{t}\zeta_1^M k_{3,M} \zeta_2^N k_{4,N}\zeta_4^Q\zeta_{3,Q}\right)\nonumber\\
    &=& -\frac{\tilde{s}\tilde{t}}{256}\bar{\lambda}_2(k_4 \cdot \Gamma)\Gamma_N\Gamma_M\lambda_3\bar{\lambda}_1(k_3 \cdot \Gamma)\Gamma^N\Gamma^M\lambda_4 \, ,
\end{eqnarray}
\begin{eqnarray}
{\cal {T}}_{1,15}&=& \left(-\frac{1}{8} \tilde{s} \bar{u}_2 \Gamma^{M'} u_{3} \bar{u}_1  \Gamma_{M'} u_4\right)\otimes\left(\frac{1}{16\cdot 2}\tilde{t}\zeta_4^Q k_{2,Q} \zeta_3^P k_{1,P}\zeta_1^M\zeta_{2,M}\right)\nonumber\\
    &=& -\frac{\tilde{s}\tilde{t}}{256}\bar{\lambda}_2\Gamma^M\Gamma^N(k_1 \cdot \Gamma)\lambda_3\bar{\lambda}_1\Gamma_M\Gamma_N(k_2 \cdot \Gamma)\lambda_4 \, .
\end{eqnarray}
Now, we list the terms of the form ${\cal {T}}_{2,i}$.
\begin{eqnarray}
{\cal {T}}_{2,1}&=&\left(\frac{1}{8} \tilde{t} \bar{u}_1 \Gamma^{M'} u_{2} \bar{u}_4\Gamma_{M'} u_3\right) \otimes \left(-\frac{1}{16\cdot 4}\tilde{s}\tilde{t}\zeta_1^M \zeta_{3,M} \zeta_4^Q \zeta_{2,Q}\right)\nonumber\\
    &=&-\frac{\tilde{s}\tilde{t}^2}{512}\bar{\lambda_1}\Gamma^M\Gamma^P\Gamma^N\lambda_2\bar{\lambda_4}\Gamma_N\Gamma_P\Gamma_M\lambda_3 \, ,
\end{eqnarray}
\begin{eqnarray}
{\cal {T}}_{2,2}&=&\left(\frac{1}{8} \tilde{t} \bar{u}_1 \Gamma^{M'} u_{2} \bar{u}_4  \Gamma_{M'} u_3\right)\otimes\left(-\frac{1}{16\cdot 4}\tilde{s}\tilde{u}\zeta_1^M \zeta_{4,M} \zeta_3^P \zeta_{2,P}\right) \nonumber\\
    &=&-\frac{\tilde{s}\tilde{t}\tilde{u}}{512}\bar{\lambda_1}\Gamma^M\Gamma^N\Gamma^P\lambda_2\bar{\lambda_4}\Gamma_M\Gamma_N\Gamma_P\lambda_3 \, ,
\end{eqnarray}
\begin{eqnarray}
{\cal {T}}_{2,3}&=&\left(\frac{1}{8} \tilde{t} \bar{u}_1 \Gamma^{M'} u_{2} \bar{u}_4  \Gamma_{M'} u_3\right)\otimes\left(-\frac{1}{16\cdot 4}\tilde{t}\tilde{u}\zeta_1^M \zeta_{2,M} \zeta_3^P \zeta_{4,P}\right) \nonumber\\
    &=&-\frac{1}{8}\tilde{t}^2\tilde{u}\bar{\lambda_1}\Gamma^M\lambda_2\bar{\lambda_4}\Gamma_M\lambda_3 \, ,
\end{eqnarray}
\begin{eqnarray}
{\cal {T}}_{2,4}&=&\left(\frac{1}{8} \tilde{t} \bar{u}_1 \Gamma^{M'} u_{2} \bar{u}_4  \Gamma_{M'} u_3\right)\otimes\left(\frac{1}{16\cdot 2}\tilde{s}\zeta_1^M k_{3,M} \zeta_4^Q k_{2,Q}\zeta_3^P\zeta_{2,P}\right)\nonumber\\
    &=&\frac{\tilde{s}\tilde{t}}{256}\bar{\lambda_1}(k_3 \cdot \Gamma)\Gamma_N\Gamma_M\lambda_2\bar{\lambda_4}(k_2 \cdot \Gamma)\Gamma^N\Gamma^M\lambda_3 \, ,
\end{eqnarray}
\begin{eqnarray}
{\cal {T}}_{2,5}&=&\left(\frac{1}{8} \tilde{t} \bar{u}_1 \Gamma^{M'} u_{2} \bar{u}_4 \Gamma_{M'} u_3\right)\otimes\left(\frac{1}{16\cdot 2}\tilde{s}\zeta_2^N k_{4,N} \zeta_3^P k_{1,P}\zeta_1^M\zeta_{4,M}\right)\nonumber\\
     &=&\frac{\tilde{s}\tilde{t}}{256}\bar{\lambda_1}\Gamma^M\Gamma^N(k_4 \cdot \Gamma)\lambda_2\bar{\lambda_4}\Gamma_M\Gamma_N(k_1 \cdot \Gamma)\lambda_3 \, ,
\end{eqnarray}
\begin{eqnarray}
{\cal {T}}_{2,6}&=&\left(\frac{1}{8} \tilde{t} \bar{u}_1 \Gamma^{M'} u_{2} \bar{u}_4  \Gamma_{M'} u_3\right)\otimes\left(\frac{1}{16\cdot 2}\tilde{s}\zeta_1^M k_{4,M} \zeta_3^P k_{2,P}\zeta_4^Q\zeta_{2,Q}\right)\nonumber\\
    &=&\frac{\tilde{s}\tilde{t}}{256}\bar{\lambda_1}(k_4 \cdot \Gamma)\Gamma_M\Gamma_N\lambda_2\bar{\lambda_4}\Gamma^N\Gamma^M(k_2 \cdot \Gamma)\lambda_3 \, ,
\end{eqnarray}
\begin{eqnarray}
{\cal {T}}_{2,7}&=&\left(\frac{1}{8} \tilde{t} \bar{u}_1 \Gamma^{M'} u_{2} \bar{u}_4  \Gamma_{M'} u_3\right)\otimes\left(\frac{1}{16\cdot 2}\tilde{s}\zeta_2^N k_{3,N} \zeta_4^Q k_{1,Q}\zeta_1^M\zeta_{3,M}\right)\nonumber\\
   &=&\frac{\tilde{s}\tilde{t}}{256}\bar{\lambda_1}\Gamma^M\Gamma^N(k_3 \cdot \Gamma)\lambda_2\bar{\lambda_4}(k_1 \cdot \Gamma)\Gamma_N\Gamma_M\lambda_3 \, ,
\end{eqnarray}
\begin{eqnarray}
{\cal {T}}_{2,8}&=&\left(\frac{1}{8} \tilde{t} \bar{u}_1 \Gamma^{M'} u_{2} \bar{u}_4  \Gamma_{M'} u_3\right)\otimes\left(\frac{1}{16\cdot 2}\tilde{u}\zeta_2^N k_{1,N} \zeta_3^P k_{4,P}\zeta_4^Q\zeta_{1,Q}\right)\nonumber\\
  &=&\frac{\tilde{t}\tilde{u}}{256}\bar{\lambda_1}\Gamma_M\Gamma_N(k_1 \cdot \Gamma)\lambda_2\bar{\lambda_4}\Gamma^M\Gamma^N(k_4 \cdot \Gamma)\lambda_3\nonumber\\
  &=& \frac{\tilde{t}\tilde{u}}{16}k_{1[N}\bar{\lambda}_1\Gamma_{M]}\lambda_2 k_4^{[N}\bar{\lambda}_4\Gamma^{M]}\lambda_3 \, ,
\end{eqnarray}
\begin{eqnarray}
{\cal {T}}_{2,9}&=&\left(\frac{1}{8} \tilde{t} \bar{u}_1 \Gamma^{M'} u_{2} \bar{u}_4  \Gamma_{M'} u_3\right)\otimes\left(\frac{1}{16\cdot 2}\tilde{u}\zeta_4^Q k_{3,Q} \zeta_1^M k_{2,M}\zeta_2^N\zeta_{3,N}\right)\nonumber\\
   &=&\frac{\tilde{t}\tilde{u}}{256}\bar{\lambda_1}(k_2 \cdot \Gamma)\Gamma^M\Gamma^N\lambda_2\bar{\lambda_4}(k_3 \cdot \Gamma)\Gamma_M\Gamma_N\lambda_3\nonumber\\
  &=& \frac{\tilde{t}\tilde{u}}{16}k_2^{[N}\bar{\lambda}_1\Gamma^{M]}\lambda_2 k_{3[N}\bar{\lambda}_4\Gamma_{M]}\lambda_3 \, ,
\end{eqnarray}
\begin{eqnarray}
{\cal {T}}_{2,10}&=&\left(\frac{1}{8} \tilde{t} \bar{u}_1 \Gamma^{M'} u_{2} \bar{u}_4  \Gamma_{M'} u_3\right)\otimes\left(\frac{1}{16\cdot 2}\tilde{u}\zeta_2^N k_{3,N} \zeta_1^M k_{4,M}\zeta_4^Q\zeta_{3,Q}\right)\nonumber\\
    &=&\frac{\tilde{t}\tilde{u}}{32}\bar{\lambda}_1(k_4 \cdot \Gamma)\Gamma^M (k_3 \cdot \Gamma)\lambda_2\bar{\lambda}_4\Gamma_M \lambda_3 \, ,
\end{eqnarray}
\begin{eqnarray}
{\cal {T}}_{2,11}&=&\left(\frac{1}{8} \tilde{t}\bar{u}_1 \Gamma^{M'} u_{2} \bar{u}_4  \Gamma_{M'} u_3\right)\otimes\left(\frac{1}{16\cdot 2}\tilde{u}\zeta_4^Q k_{1,Q} \zeta_3^P k_{2,P}\zeta_2^N\zeta_{1,N}\right)\nonumber\\
    &=& \frac{\tilde{t}\tilde{u}}{32}\bar{\lambda}_1 \Gamma_M\lambda_2\bar{\lambda}_4(k_1 \cdot \Gamma)\Gamma^M (k_2 \cdot \Gamma)\lambda_3 \, ,
\end{eqnarray}
\begin{eqnarray}
{\cal {T}}_{2,12}&=&\left(\frac{1}{8} \tilde{t} \bar{u}_1 \Gamma^{M'} u_{2} \bar{u}_4  \Gamma_{M'} u_3\right)\otimes\left(\frac{1}{16\cdot 2}\tilde{t}\zeta_1^M k_{2,M} \zeta_3^P k_{4,P}\zeta_4^Q\zeta_{2,Q}\right)\nonumber\\
   &=&\frac{\tilde{t}^2}{256}\bar{\lambda_1}(k_2 \cdot \Gamma)\Gamma^M\Gamma^N\lambda_2\bar{\lambda_4}\Gamma_N\Gamma_M(k_4 \cdot \Gamma)\lambda_3\nonumber\\
  &=& \frac{\tilde{t}^2}{16}k_2^{[N}\bar{\lambda}_1\Gamma^{M]}\lambda_2 k_{4[N}\bar{\lambda}_4\Gamma_{M]}\lambda_3 \, ,
\end{eqnarray}
\begin{eqnarray}
{\cal {T}}_{2,13}&=&\left(\frac{1}{8} \tilde{t} \bar{u}_1 \Gamma^{M'} u_{2} \bar{u}_4  \Gamma_{M'} u_3\right)\otimes\left(\frac{1}{16\cdot 2}\tilde{t}\zeta_4^Q k_{3,Q} \zeta_2^N k_{1,N}\zeta_1^M\zeta_{3,M}\right)\nonumber\\
  &=&\frac{\tilde{t}^2}{256}\bar{\lambda_1}\Gamma^M\Gamma^N(k_1 \cdot \Gamma)\lambda_2\bar{\lambda_4}(k_3 \cdot \Gamma)\Gamma_N\Gamma_M\lambda_3\nonumber\\
  &=& \frac{\tilde{t}^2}{16}k_1^{[N}\bar{\lambda}_1\Gamma^{M]}\lambda_2 k_{3[N}\bar{\lambda}_4\Gamma_{M]}\lambda_3  \, ,
\end{eqnarray}
\begin{eqnarray}
{\cal {T}}_{2,14}&=& \left(\frac{1}{8} \tilde{t} \bar{u}_1 \Gamma^{M'} u_{2} \bar{u}_4  \Gamma_{M'} u_3\right)\otimes\left(\frac{1}{16\cdot 2}\tilde{t}\zeta_1^M k_{3,M} \zeta_2^Q k_{4,Q}\zeta_4^N\zeta_{3,N}\right)\nonumber\\
 &=&\frac{\tilde{t}^2}{32}\bar{\lambda}_1(k_3 \cdot \Gamma)\Gamma^M (k_4 \cdot \Gamma)\lambda_2\bar{\lambda}_4\Gamma_M \lambda_3  \, ,
\end{eqnarray}
\begin{eqnarray}
{\cal {T}}_{2,15}&=& \left(\frac{1}{8} \tilde{t} \bar{u}_1 \Gamma^{M'} u_{2} \bar{u}_4  \Gamma_{M'} u_3\right)\otimes\left(\frac{1}{16\cdot 2}\tilde{t}\zeta_4^Q k_{2,Q} \zeta_3^P k_{1,P}\zeta_1^M\zeta_{2,M}\right)\nonumber\\
  &=& \frac{\tilde{t}^2}{32}\bar{\lambda}_1 \Gamma_M\lambda_2\bar{\lambda}_4(k_2 \cdot \Gamma)\Gamma^M (k_1 \cdot \Gamma)\lambda_3   \, .
\end{eqnarray}

\section{Tensor-product terms after using the kinematics}

In this appendix we show explicitly the form of the terms ${\cal {T}}_{1, i}$ and ${\cal {T}}_{2, i}$, considering  the kinematics (\ref{kinematics}), for $i \ge 4$, 
\begin{eqnarray}
{\cal {T}}_{1, 4}&=& -\frac{\tilde{s}^2}{32}[\frac{\tilde{u}}{2}\bar{\lambda}_2\Gamma^0\lambda_3\bar{\lambda}_1 \Gamma_0\lambda_4-\frac{\tilde{u}}{2}\bar{\lambda}_2\Gamma^0\lambda_3\bar{\lambda}_1 \Gamma_1\lambda_4 \nonumber \\
&+&\frac{\tilde{s}}{4}\sin\theta\bar{\lambda}_2\Gamma^0\lambda_3\bar{\lambda}_1 \Gamma_2\lambda_4-\frac{\tilde{u}}{2}\bar{\lambda}_2\Gamma^1\lambda_3\bar{\lambda}_1\Gamma_1\lambda_4-\frac{\tilde{t}}{2}\bar{\lambda}_2\Gamma^{2}\lambda_3\bar{\lambda}_1\Gamma_{2}\lambda_4\nonumber\\
&-&\frac{\tilde{t}}{2}\bar{\lambda}_2\Gamma^{i_{>2}}\lambda_3\bar{\lambda}_1\Gamma_{i_{>2}}\lambda_4-\frac{\tilde{s}}{4}\sin\theta\bar{\lambda}_2\Gamma^0\lambda_3\bar{\lambda}_1 \Gamma_0\Gamma_1\Gamma_2\lambda_4+\frac{\tilde{u}}{2}\bar{\lambda}_2\Gamma^1\lambda_3\bar{\lambda}_1\Gamma_0\lambda_4\nonumber\\
&+&\frac{\tilde{t}}{2}\bar{\lambda}_2\Gamma^{2}\lambda_3\bar{\lambda}_1 \Gamma_0\Gamma_1\Gamma_{2}\lambda_4+\frac{\tilde{t}}{2}\bar{\lambda}_2\Gamma^{i_{>2}}\lambda_3\bar{\lambda}_1 \Gamma_0\Gamma_1\Gamma_{i_{>2}}\lambda_4-\frac{\tilde{s}}{4}\sin\theta\bar{\lambda}_2\Gamma^{2}\lambda_3\bar{\lambda}_1\Gamma_0\lambda_4\nonumber\\
&+&\frac{\tilde{s}}{4}\sin\theta\bar{\lambda}_2\Gamma^{1}\lambda_3\bar{\lambda}_1\Gamma_0\Gamma_1\Gamma_{2}\lambda_4+\frac{\tilde{s}}{4}\sin\theta\bar{\lambda}_2\Gamma^{i_{>2}}\lambda_3\bar{\lambda}_1\Gamma_0\Gamma_2\Gamma_{i_{>2}}\lambda_4+\frac{\tilde{s}}{4}\sin\theta\bar{\lambda}_2\Gamma^1\lambda_3\bar{\lambda}_1\Gamma_2\lambda_4\nonumber\\
&+&\frac{\tilde{s}}{4}\sin\theta\bar{\lambda}_2\Gamma^2\lambda_3\bar{\lambda}_1\Gamma_1\lambda_4-\frac{\tilde{s}}{4}\sin\theta\bar{\lambda}_2\Gamma^{i_{>2}}\lambda_3\bar{\lambda}_1\Gamma_1\Gamma_2\Gamma_{i_{>2}}\lambda_4] \, ,
\end{eqnarray}
\begin{eqnarray}
{\cal {T}}_{1,5}&=& -\frac{\tilde{s}^2}{32}[\frac{\tilde{u}}{2}\bar{\lambda}_2\Gamma^0\lambda_3\bar{\lambda}_1\Gamma_0\lambda_4-\frac{\tilde{u}}{2}\bar{\lambda}_2 \Gamma^1\lambda_3\bar{\lambda}_1\Gamma_0\lambda_4 \nonumber \\
&+&\frac{\tilde{s}}{4}\sin\theta\bar{\lambda}_2\Gamma^2\lambda_3\bar{\lambda}_1\Gamma_0\lambda_4-\frac{\tilde{s}}{4}\sin\theta\bar{\lambda}_2\Gamma^0\Gamma^1 \Gamma^2\lambda_3\bar{\lambda}_1\Gamma_0\lambda_4\nonumber\\
&-&\frac{\tilde{u}}{2}\bar{\lambda}_2\Gamma^1 \lambda_3\bar{\lambda}_1\Gamma_1\lambda_4-\frac{\tilde{t}}{2}\bar{\lambda}_2\Gamma^2 \lambda_3\bar{\lambda}_1\Gamma_2\lambda_4-\frac{\tilde{t}}{2}\bar{\lambda}_2\Gamma^{i_{>2}} \lambda_3\bar{\lambda}_1\Gamma_{i_{>2}}\lambda_4\nonumber\\
&+&\frac{\tilde{u}}{2}\bar{\lambda}_2\Gamma^0\lambda_3\bar{\lambda}_1\Gamma_1\lambda_4+\frac{\tilde{t}}{2}\bar{\lambda}_2\Gamma^0\Gamma^1 \Gamma^2\lambda_3\bar{\lambda}_1\Gamma_2\lambda_4+\frac{\tilde{t}}{2}\bar{\lambda}_2\Gamma^0\Gamma^1\Gamma^{i_{>2}} \lambda_3\bar{\lambda}_1\Gamma_{i_{>2}}\lambda_4\nonumber\\
&-&\frac{\tilde{s}}{4}\sin\theta\bar{\lambda}_2\Gamma^0\Gamma^1 \Gamma^2\lambda_3\bar{\lambda}_1\Gamma_1\lambda_4-\frac{\tilde{s}}{4}\sin\theta\bar{\lambda}_2 \Gamma^0\lambda_3\bar{\lambda}_1\Gamma_2\lambda_4+\frac{\tilde{s}}{4}\sin\theta\bar{\lambda}_2\Gamma^0\Gamma^2\Gamma^{i_{>2}}  \lambda_3\bar{\lambda}_1\Gamma_{i_{>2}} \lambda_4\nonumber\\ 
&+&\frac{\tilde{s}}{4}\sin\theta\bar{\lambda}_2\Gamma^2\lambda_3\bar{\lambda}_1\Gamma_1\lambda_4+\frac{\tilde{s}}{4}\sin\theta\bar{\lambda}_2 \Gamma^1\lambda_3\bar{\lambda}_1\Gamma_2\lambda_4-\frac{\tilde{s}}{4}\sin\theta\bar{\lambda}_2\Gamma^1\Gamma^2\Gamma^{i_{>2}} \lambda_3\bar{\lambda}_1\Gamma_{i_{>2}}\lambda_4] 
\end{eqnarray}
\begin{eqnarray}
{\cal {T}}_{1,6}&=& -\frac{\tilde{s}^3}{128}[\bar{\lambda}_2\Gamma^1\lambda_3\bar{\lambda}_1\Gamma_1\lambda_4+\cos\theta\bar{\lambda}_2\Gamma^1\lambda_3\bar{\lambda}_1\Gamma_0\lambda_4+\bar{\lambda}_2\Gamma^0\lambda_3\bar{\lambda}_1\Gamma_1\lambda_4\nonumber\\
&+&\cos\theta\bar{\lambda}_2\Gamma^0\lambda_3\bar{\lambda}_1\Gamma_0\lambda_4+\sin\theta\bar{\lambda}_2\Gamma^2\lambda_3\bar{\lambda}_1\Gamma_0\lambda_4-\sin\theta\bar{\lambda}_2\Gamma^2\lambda_3\bar{\lambda}_1\Gamma_1\lambda_4 \nonumber \\
&-&2\frac{\tilde{u}}{\tilde{s}}\bar{\lambda}_2\Gamma^{i_{>1}}\lambda_3\bar{\lambda}_1\Gamma_{i_{>1}}\lambda_4] \, ,
\end{eqnarray}
\begin{eqnarray}
{\cal {T}}_{1,7}&=& -\frac{\tilde{s}^3}{128}[\bar{\lambda}_2\Gamma^1\lambda_3\bar{\lambda}_1\Gamma_1\lambda_4+\bar{\lambda}_2\Gamma^1\lambda_3\bar{\lambda}_1\Gamma_0\lambda_4+\cos\theta\bar{\lambda}_2\Gamma^0\lambda_3\bar{\lambda}_1\Gamma_1\lambda_4 \nonumber \\
&+&\cos\theta\bar{\lambda}_2\Gamma^0\lambda_3\bar{\lambda}_1\Gamma_0\lambda_4
+\sin\theta\bar{\lambda}_2\Gamma^0\lambda_3\bar{\lambda}_1\Gamma_2\lambda_4-\sin\theta\bar{\lambda}_2\Gamma^1\lambda_3\bar{\lambda}_1\Gamma_2\lambda_4 \nonumber \\
&-&2\frac{\tilde{u}}{\tilde{s}}\bar{\lambda}_2\Gamma^{i_{>1}}\lambda_3 \bar{\lambda}_1\Gamma_{i_{>1}}\lambda_4] \, ,
\end{eqnarray}
\begin{eqnarray}
{\cal {T}}_{1,8} &=& -\frac{\tilde{s}\tilde{u}}{32}[\frac{\tilde{u}}{2}\bar{\lambda}_2\Gamma^0\lambda_3\bar{\lambda}_1\Gamma_0\lambda_4-\frac{\tilde{u}}{2}\bar{\lambda}_2 \Gamma^1\lambda_3\bar{\lambda}_1\Gamma_0\lambda_4 
+\frac{\tilde{s}}{4}\sin\theta\bar{\lambda}_2\Gamma^2\lambda_3\bar{\lambda}_1\Gamma_0\lambda_4 \nonumber \\
&+&\frac{\tilde{s}}{4}\sin\theta\bar{\lambda}_2\Gamma^0\Gamma^1 \Gamma^2\lambda_3\bar{\lambda}_1\Gamma_0\lambda_4
-\frac{\tilde{u}}{2}\bar{\lambda}_2\Gamma^1 \lambda_3\bar{\lambda}_1\Gamma_1\lambda_4
-\frac{\tilde{t}}{2}\bar{\lambda}_2\Gamma^2 \lambda_3\bar{\lambda}_1\Gamma_2\lambda_4 \nonumber \\
&-&\frac{\tilde{t}}{2}\bar{\lambda}_2\Gamma^{i_{>2}} \lambda_3\bar{\lambda}_1\Gamma_{i_{>2}}\lambda_4+\frac{\tilde{u}}{2}\bar{\lambda}_2\Gamma^0\lambda_3\bar{\lambda}_1\Gamma_1\lambda_4-\frac{\tilde{t}}{2}\bar{\lambda}_2\Gamma^0\Gamma^1 \Gamma^2\lambda_3\bar{\lambda}_1\Gamma_2\lambda_4 \nonumber \\
&-&\frac{\tilde{t}}{2}\bar{\lambda}_2\Gamma^0\Gamma^1\Gamma^{i_{>2}} \lambda_3\bar{\lambda}_1\Gamma_{i_{>2}}\lambda_4-\frac{\tilde{s}}{4}\sin\theta\bar{\lambda}_2\Gamma^0\Gamma^2\Gamma^{i_{>2}} \lambda_3\bar{\lambda}_1\Gamma_{i_{>2}}\lambda_4\nonumber\\
&+&\frac{\tilde{s}}{4}\sin\theta\bar{\lambda}_2\Gamma^0\Gamma^1\Gamma^{2} \lambda_3\bar{\lambda}_1\Gamma_{1}\lambda_4-\frac{\tilde{s}}{4}\sin\theta\bar{\lambda}_2\Gamma^0 \lambda_3\bar{\lambda}_1\Gamma_{2}\lambda_4 \nonumber \\
&+&\frac{\tilde{s}}{4}\sin\theta\bar{\lambda}_2\Gamma^1\Gamma^2\Gamma^{i_{>2}} \lambda_3\bar{\lambda}_1\Gamma_{i_{>2}}\lambda_4+\frac{\tilde{s}}{4}\sin\theta\bar{\lambda}_2\Gamma^2\lambda_3\bar{\lambda}_1\Gamma_{1}\lambda_4\nonumber\\
&+&\frac{\tilde{s}}{4}\sin\theta\bar{\lambda}_2\Gamma^1 \lambda_3\bar{\lambda}_1\Gamma_{2}\lambda_4] \, ,
\end{eqnarray}
\begin{eqnarray}
{\cal {T}}_{1,9}&=& -\frac{\tilde{s}\tilde{u}}{32}[\frac{\tilde{u}}{2}\bar{\lambda}_2\Gamma^0\lambda_3\bar{\lambda}_1 \Gamma_0\lambda_4-\frac{\tilde{u}}{2}\bar{\lambda}_2\Gamma^0\lambda_3\bar{\lambda}_1 \Gamma_1\lambda_4+\frac{\tilde{s}}{4}\sin\theta\bar{\lambda}_2\Gamma^0\lambda_3\bar{\lambda}_1\Gamma_2\lambda_4\nonumber\\
&+&\frac{\tilde{s}}{4}\sin\theta\bar{\lambda}_2\Gamma^0\lambda_3\bar{\lambda}_1\Gamma_0\Gamma_1\Gamma_2\lambda_4-\frac{\tilde{u}}{2}\bar{\lambda}_2\Gamma^1\lambda_3\bar{\lambda}_1\Gamma_1\lambda_4-\frac{\tilde{t}}{2}\bar{\lambda}_2\Gamma^{2}\lambda_3\bar{\lambda}_1\Gamma_{2}\lambda_4\nonumber\\
&-&\frac{\tilde{t}}{2}\bar{\lambda}_2\Gamma^{i_{>2}}\lambda_3\bar{\lambda}_1\Gamma_{i_{>2}}\lambda_4-\frac{\tilde{t}}{2}\bar{\lambda}_2\Gamma^{i_{>2}}\lambda_3\bar{\lambda}_1 \Gamma_0\Gamma_1\Gamma_{i_{>2}}\lambda_4+\frac{\tilde{u}}{2}\bar{\lambda}_2\Gamma^1\lambda_3\bar{\lambda}_1\Gamma_0\lambda_4 \nonumber \\
&-&\frac{\tilde{t}}{2}\bar{\lambda}_2\Gamma^{2}\lambda_3\bar{\lambda}_1\Gamma_0\Gamma_1\Gamma_{2}\lambda_4
-\frac{\tilde{s}}{4}\sin\theta\bar{\lambda}_2\Gamma^{i_{>2}}\lambda_3\bar{\lambda}_1 \Gamma_0\Gamma_2\Gamma_{i_{>2}}\lambda_4 \nonumber \\
&+&\frac{\tilde{s}}{4}\sin\theta\bar{\lambda}_2\Gamma^{1}\lambda_3\bar{\lambda}_1 \Gamma_0\Gamma_1\Gamma_{2}\lambda_4-\frac{\tilde{s}}{4}\sin\theta\bar{\lambda}_2\Gamma^{2}\lambda_3\bar{\lambda}_1\Gamma_0\lambda_4 \nonumber \\
&+&\frac{\tilde{s}}{4}\sin\theta\bar{\lambda}_2\Gamma^{i_{>2}}\lambda_3\bar{\lambda}_1\Gamma_1\Gamma_2\Gamma_{i_{>2}}\lambda_4+\frac{\tilde{s}}{4}\sin\theta\bar{\lambda}_2\Gamma^1\lambda_3\bar{\lambda}_1\Gamma_2\lambda_4 \nonumber \\
&+&\frac{\tilde{s}}{4}\sin\theta\bar{\lambda}_2\Gamma^2\lambda_3\bar{\lambda}_1\Gamma_1\lambda_4] \, ,
\end{eqnarray}
\begin{eqnarray}
{\cal {T}}_{1,10}&=&-\frac{\tilde{s}^2\tilde{u}}{128}[-\bar{\lambda}_2\Gamma^{1}\lambda_3 \bar{\lambda}_1\Gamma_{1}\lambda_4-\bar{\lambda}_2\Gamma^{1}\lambda_3 \bar{\lambda}_1\Gamma_{0}\lambda_4-\bar{\lambda}_2\Gamma^{0}\lambda_3 \bar{\lambda}_1\Gamma_{1}\lambda_4\nonumber\\
&-&\bar{\lambda}_2\Gamma^{0}\lambda_3 \bar{\lambda}_1\Gamma_{0}\lambda_4
-2\bar{\lambda}_2\Gamma^{i_{>1}}\lambda_3\bar{\lambda}_1\Gamma_{i_{>1}}\lambda_4] \, ,
\end{eqnarray}
\begin{eqnarray}
{\cal {T}}_{1,11} &=&-\frac{\tilde{s}^2\tilde{u}}{128}[-\bar{\lambda}_2\Gamma^{1}\lambda_3 \bar{\lambda}_1\Gamma_{1}\lambda_4-\bar{\lambda}_2\Gamma^{1}\lambda_3 \bar{\lambda}_1\Gamma_{0}\lambda_4-\bar{\lambda}_2\Gamma^{0}\lambda_3 \bar{\lambda}_1\Gamma_{1}\lambda_4\nonumber\\
&-&\bar{\lambda}_2\Gamma^{0}\lambda_3 \bar{\lambda}_1\Gamma_{0}\lambda_4
-2\bar{\lambda}_2\Gamma^{i_{>1}}\lambda_3\bar{\lambda}_1\Gamma_{i_{>1}}\lambda_4] \, ,
\end{eqnarray}
\begin{eqnarray}
{\cal {T}}_{1,12} &=& -\frac{\tilde{s}^2\tilde{t}}{1024}[\bar{\lambda}_2\Gamma^M\Gamma^N\Gamma^0\lambda_3\bar{\lambda}_1\Gamma_{0}\Gamma_N\Gamma_M\lambda_4-\bar{\lambda}_2\Gamma^0\Gamma^M\Gamma^N\lambda_3\bar{\lambda}_1\Gamma_{1}\Gamma_N\Gamma_M\lambda_4\nonumber\\
&+&\cos\theta\bar{\lambda}_2\Gamma^M\Gamma^N\Gamma^1\lambda_3\bar{\lambda}_1\Gamma_{1}\Gamma_N\Gamma_M\lambda_4-\cos\theta\bar{\lambda}_2\Gamma^M\Gamma^N\Gamma^1\lambda_3\bar{\lambda}_1\Gamma_{0}\Gamma_N\Gamma_M\lambda_4\nonumber\\
&-&\sin\theta\bar{\lambda}_2\Gamma^M\Gamma^N\Gamma^2\lambda_3\bar{\lambda}_1\Gamma_{0}\Gamma_N\Gamma_M\lambda_4+\sin\theta\bar{\lambda}_2\Gamma^M\Gamma^N\Gamma^2\lambda_3\bar{\lambda}_1\Gamma_{1}\Gamma_N\Gamma_M\lambda_4] \, ,
\end{eqnarray}
\begin{eqnarray}
{\cal {T}}_{1,13}&=&  -\frac{\tilde{s}^2\tilde{t}}{1024}[\bar{\lambda}_2\Gamma^0\Gamma^N\Gamma^M\lambda_3\bar{\lambda}_1\Gamma_{M}\Gamma_N\Gamma_0\lambda_4-\cos\theta\bar{\lambda}_2\Gamma^0\Gamma^N\Gamma^M\lambda_3\bar{\lambda}_1\Gamma_{M}\Gamma_N\Gamma_1\lambda_4\nonumber\\
&-&\sin\theta\bar{\lambda}_2\Gamma^0\Gamma^N\Gamma^M\lambda_3\bar{\lambda}_1\Gamma_{M}\Gamma_N\Gamma_2\lambda_4-\bar{\lambda}_2\Gamma^1\Gamma^N\Gamma^M\lambda_3\bar{\lambda}_1\Gamma_{M}\Gamma_N\Gamma_0\lambda_4\nonumber\\
&+&\cos\theta\bar{\lambda}_2\Gamma^1\Gamma^N\Gamma^M\lambda_3\bar{\lambda}_1\Gamma_{M}\Gamma_N\Gamma_1\lambda_4+\sin\theta\bar{\lambda}_2\Gamma^1\Gamma^N\Gamma^M\lambda_3\bar{\lambda}_1\Gamma_{M}\Gamma_N\Gamma_2\lambda_4] \, ,
\end{eqnarray}
\begin{eqnarray}
{\cal {T}}_{1,14}&=&-\frac{\tilde{s}^2\tilde{t}}{1024}[-\bar{\lambda}_2\Gamma^0\Gamma^N\Gamma^M\lambda_3\bar{\lambda}_1\Gamma_{0}\Gamma_N\Gamma_M\lambda_4-cos^2\theta\bar{\lambda}_2\Gamma^1\Gamma^N\Gamma^M\lambda_3\bar{\lambda}_1\Gamma_{1}\Gamma_N\Gamma_M\lambda_4\nonumber\\
&+&\cos\theta\bar{\lambda}_2\Gamma^0\Gamma^N\Gamma^M\lambda_3\bar{\lambda}_1\Gamma_{1}\Gamma_N\Gamma_M\lambda_4+\sin\theta\bar{\lambda}_2\Gamma^0\Gamma^N\Gamma^M\lambda_3\bar{\lambda}_1\Gamma_{2}\Gamma_N\Gamma_M\lambda_4\nonumber\\
&+&\cos\theta\bar{\lambda}_2\Gamma^1\Gamma^N\Gamma^M\lambda_3\bar{\lambda}_1\Gamma_{0}\Gamma_N\Gamma_M\lambda_4-\sin\theta \cos\theta\bar{\lambda}_2\Gamma^1\Gamma^N\Gamma^M\lambda_3\bar{\lambda}_1\Gamma_{2}\Gamma_N\Gamma_M\lambda_4\nonumber\\
&+&\sin\theta \bar{\lambda}_2\Gamma^2\Gamma^N\Gamma^M\lambda_3\bar{\lambda}_1\Gamma_{0}\Gamma_N\Gamma_M\lambda_4-\sin\theta \cos\theta\bar{\lambda}_2\Gamma^2\Gamma^N\Gamma^M\lambda_3\bar{\lambda}_1\Gamma_{1}\Gamma_N\Gamma_M\lambda_4\nonumber\\
&-& sin^2\theta
\bar{\lambda}_2\Gamma^2\Gamma^N\Gamma^M\lambda_3\bar{\lambda}_1\Gamma_{2}\Gamma_N\Gamma_M\lambda_4] \, ,
\end{eqnarray}
\begin{eqnarray}
{\cal {T}}_{1,15}
&=&-\frac{\tilde{s}^2\tilde{t}}{1024}[-\bar{\lambda}_2\Gamma^M\Gamma^N\Gamma^0\lambda_3\bar{\lambda}_1\Gamma_{M}\Gamma_N\Gamma_0\lambda_4-\bar{\lambda}_2\Gamma^M\Gamma^N\Gamma^1\lambda_3\bar{\lambda}_1\Gamma_{M}\Gamma_N\Gamma_1\lambda_4\nonumber\\
&+&\bar{\lambda}_2\Gamma^M\Gamma^N\Gamma^0\lambda_3\bar{\lambda}_1\Gamma_{M}\Gamma_N\Gamma_1\lambda_4+\bar{\lambda}_2\Gamma^M\Gamma^N\Gamma^1\lambda_3\bar{\lambda}_1\Gamma_{M}\Gamma_N\Gamma_0\lambda_4] \, .
\end{eqnarray}

Also, we have obtained the following terms:
\begin{eqnarray}
{\cal {T}}_{2,4}
 &=&\frac{\tilde{s}^2\tilde{t}}{1024}[\bar{\lambda_1}\Gamma^0\Gamma^N\Gamma^M\lambda_2\bar{\lambda_4} \Gamma_{0}\Gamma_N\Gamma_M\lambda_3-\bar{\lambda_1}\Gamma^0\Gamma^N\Gamma^M\lambda_2\bar{\lambda_4} \Gamma_{1}\Gamma_N\Gamma_M\lambda_3\nonumber\\
 \nonumber\\
  &+&\cos\theta\bar{\lambda_1}\Gamma^1\Gamma^N\Gamma^M\lambda_2\bar{\lambda_4} \Gamma_{0}\Gamma_N\Gamma_M\lambda_3-\cos\theta\bar{\lambda_1}\Gamma^1\Gamma^N\Gamma^M\lambda_2\bar{\lambda_4} \Gamma_{1}\Gamma_N\Gamma_M\lambda_3\nonumber\\
  &+&\sin\theta\bar{\lambda_1}\Gamma^2\Gamma^N\Gamma^M\lambda_2\bar{\lambda_4} \Gamma_{0}\Gamma_N\Gamma_M\lambda_3-\sin\theta\bar{\lambda_1}\Gamma^2\Gamma^N\Gamma^M\lambda_2\bar{\lambda_4} \Gamma_{1}\Gamma_N\Gamma_M\lambda_3]
 ]\, ,
\end{eqnarray}
\begin{eqnarray}
{\cal {T}}_{2,5}
  &=&\frac{\tilde{s}^2\tilde{t}}{1024}[\bar{\lambda_1}\Gamma^M\Gamma^N\Gamma^0\lambda_2\bar{\lambda_4} \Gamma_M\Gamma_N\Gamma_{0}\lambda_3-\cos\theta\bar{\lambda_1}\Gamma^M\Gamma^N\Gamma^1\lambda_2\bar{\lambda_4} \Gamma_M\Gamma_N\Gamma_{0}\lambda_3\nonumber\\
  &-&\sin\theta\bar{\lambda_1}\Gamma^M\Gamma^N\Gamma^2\lambda_2\bar{\lambda_4} \Gamma_M\Gamma_N\Gamma_{0}\lambda_3+\bar{\lambda_1}\Gamma^M\Gamma^N\Gamma^{0}\lambda_2\bar{\lambda_4} \Gamma_M\Gamma_N\Gamma_1\lambda_3\nonumber\\
  &-&\cos\theta\bar{\lambda_1}\Gamma^M\Gamma^N\Gamma^{1}\lambda_2\bar{\lambda_4} \Gamma_M\Gamma_N\Gamma_1\lambda_3-\sin\theta\bar{\lambda_1}\Gamma^M\Gamma^N\Gamma^{2}\lambda_2\bar{\lambda_4} \Gamma_M\Gamma_N\Gamma_1\lambda_3]\, ,
\end{eqnarray}
\begin{eqnarray}
{\cal {T}}_{2,6}
    &=& \frac{\tilde{s}^2\tilde{t}}{1024}[\bar{\lambda_1}\Gamma^{M}\Gamma^N\Gamma^0\lambda_2\bar{\lambda_4} \Gamma_{0}\Gamma_N\Gamma_M\lambda_3+\cos\theta\bar{\lambda_1}\Gamma^{M}\Gamma^N\Gamma^1\lambda_2\bar{\lambda_4} \Gamma_{0}\Gamma_N\Gamma_M\lambda_3 \nonumber \\
&+&\sin\theta\bar{\lambda_1}\Gamma^{M}\Gamma^N\Gamma^2\lambda_2\bar{\lambda_4} \Gamma_{0}\Gamma_N\Gamma_M\lambda_3+\bar{\lambda_1}\Gamma^{M}\Gamma^N\Gamma^0\lambda_2\bar{\lambda_4} \Gamma_{1}\Gamma_N\Gamma_M\lambda_3 \nonumber \\
&+&\cos\theta\bar{\lambda_1}\Gamma^{M}\Gamma^N\Gamma^1\lambda_2\bar{\lambda_4}\Gamma_{1}\Gamma_N\Gamma_M\lambda_3+\sin\theta\bar{\lambda_1}\Gamma^{M}\Gamma^N\Gamma^2\lambda_2\bar{\lambda_4} \Gamma_{1}\Gamma_N\Gamma_M\lambda_3] \, ,
\end{eqnarray}
\begin{eqnarray}
{\cal {T}}_{2,7}
    &=& \frac{\tilde{s}^2\tilde{t}}{1024}[\bar{\lambda_1}\Gamma^{M}\Gamma^N\Gamma^0\lambda_2\bar{\lambda_4} \Gamma_{0}\Gamma_N\Gamma_M\lambda_3+\cos\theta\bar{\lambda_1}\Gamma^{M}\Gamma^N\Gamma^1\lambda_2\bar{\lambda_4} \Gamma_{0}\Gamma_N\Gamma_M\lambda_3 ¿\nonumber \\
&+& \sin\theta\bar{\lambda_1}\Gamma^{M}\Gamma^N\Gamma^2\lambda_2\bar{\lambda_4} \Gamma_{0}\Gamma_N\Gamma_M\lambda_3+\bar{\lambda_1}\Gamma^{M}\Gamma^N\Gamma^0\lambda_2\bar{\lambda_4} \Gamma_{1}\Gamma_N\Gamma_M\lambda_3 \nonumber \\
&+&\cos\theta\bar{\lambda_1}\Gamma^{M}\Gamma^N\Gamma^1\lambda_2\bar{\lambda_4}\Gamma_{1}\Gamma_N\Gamma_M\lambda_3+\sin\theta\bar{\lambda_1}\Gamma^{M}\Gamma^N\Gamma^2\lambda_2\bar{\lambda_4} \Gamma_{1}\Gamma_N\Gamma_M\lambda_3]\, ,
\end{eqnarray}
\begin{eqnarray}
{\cal {T}}_{2,8}
&=&\frac{\tilde{s}\tilde{t}\tilde{u}}{128}[\bar{\lambda_1}\Gamma^1\lambda_2\bar{\lambda_4}\Gamma_1\lambda_3-\cos\theta\bar{\lambda_1}\Gamma^0\lambda_2\bar{\lambda_4}\Gamma_0\lambda_3+\cos\theta\bar{\lambda_1}\Gamma^1\lambda_2\bar{\lambda_4}\Gamma_0\lambda_3\nonumber\\
&-&\bar{\lambda_1}\Gamma^0\lambda_2\bar{\lambda_4}\Gamma_1\lambda_3+\sin\theta\bar{\lambda_1}\Gamma^2\lambda_2\bar{\lambda_4}\Gamma_0\lambda_3+\sin\theta\bar{\lambda_1}\Gamma^2\lambda_2\bar{\lambda_4}\Gamma_1\lambda_3 \nonumber \\
&-&2\frac{\tilde{t}}{\tilde{s}}\bar{\lambda_1}\Gamma^{i_{>1}}\lambda_2\bar{\lambda_4}\Gamma_{i_{>1}}\lambda_3]\, ,
\end{eqnarray} 
\begin{eqnarray}
{\cal {T}}_{2,9}
&=&\frac{\tilde{s}\tilde{t}\tilde{u}}{128}[\bar{\lambda_1}\Gamma^1\lambda_2\bar{\lambda_4}\Gamma_1\lambda_3-\cos\theta\bar{\lambda_1}\Gamma^1\lambda_2\bar{\lambda_4}\Gamma_0\lambda_3+\bar{\lambda_1}\Gamma^0\lambda_2\bar{\lambda_4}\Gamma_1\lambda_3 \nonumber \\
&-&\cos\theta\bar{\lambda_1}\Gamma^0\lambda_2\bar{\lambda_4}\Gamma_0\lambda_3-\sin\theta\bar{\lambda_1}\Gamma^2\lambda_2\bar{\lambda_4}\Gamma_0\lambda_3-2\frac{\tilde{t}}{\tilde{s}}\bar{\lambda_1}\Gamma^{i_{>1}}\lambda_2\bar{\lambda_4}\Gamma_{i_{>1}}\lambda_3
\nonumber \\
&+&\sin\theta\bar{\lambda_1}\Gamma^2\lambda_2\bar{\lambda_4}\Gamma_1\lambda_3] \, ,
\end{eqnarray}
\begin{eqnarray}
{\cal {T}}_{2,10}
&=& \frac{\tilde{s}\tilde{t}\tilde{u}}{64}[
-\bar{\lambda}_1\Gamma^{i_>2} \lambda_2\bar{\lambda}_4\Gamma_{i_>2}\lambda_3-\cos\theta\bar{\lambda}_1\Gamma^0\Gamma^1\Gamma^{i_>2} \lambda_2\bar{\lambda}_4\Gamma_{i_>2}\lambda_3 \nonumber \\
&-&\sin\theta\bar{\lambda}_1\Gamma^0\Gamma^2\Gamma^{i_>2} \lambda_2\bar{\lambda}_4\Gamma_{i_>2}\lambda_3
-\cos\theta\bar{\lambda}_1\Gamma^0\Gamma^1 \Gamma^2\lambda_2\bar{\lambda}_4\Gamma_2\lambda_3 \nonumber \\
&-&\cos^2\theta\bar{\lambda}_1\Gamma^2 \lambda_2\bar{\lambda}_4\Gamma_2\lambda_3+\sin\theta\bar{\lambda}_1\Gamma^0\Gamma^1\Gamma^{2} \lambda_2\bar{\lambda}_4\Gamma_{1}\lambda_3 \nonumber \\
&-&\sin^2\theta\bar{\lambda}_1\Gamma^1 \lambda_2\bar{\lambda}_4\Gamma_{1}\lambda_3+\sin\theta \cos\theta\bar{\lambda}_1\Gamma^1\lambda_2\bar{\lambda}_4\Gamma_{2}\lambda_3] \, ,
\end{eqnarray}
\begin{eqnarray}
{\cal {T}}_{2,11}
&=& -\frac{\tilde{s}\tilde{t}\tilde{u}}{64}[
\bar{\lambda}_1\Gamma^{i_{>1}}\lambda_2\bar{\lambda}_4 \Gamma_{i_{>1}}\lambda_3-\bar{\lambda}_1\Gamma^{i_{>2}}\lambda_2\bar{\lambda}_4\Gamma_0\Gamma_1\Gamma_{i_{>2}}\lambda_3  \nonumber \\
&-&\bar{\lambda}_1\Gamma^2\lambda_2\bar{\lambda}_4\Gamma_0\Gamma_1\Gamma_2\lambda_3]
\end{eqnarray}
\begin{eqnarray}
{\cal {T}}_{2,12}  
&=& \frac{\tilde{s}\tilde{t}^2}{128}[\bar{\lambda}_1\Gamma^1\lambda_2\bar{\lambda}_4\Gamma_1\lambda_3+\cos\theta\bar{\lambda}_1\Gamma^1\lambda_2\bar{\lambda}_4\Gamma_0\lambda_3+\bar{\lambda}_1\Gamma^0\lambda_2\bar{\lambda}_4\Gamma_1\lambda_3+\sin\theta\bar{\lambda}_1\Gamma^2\lambda_2\bar{\lambda}_4\Gamma_0\lambda_3\nonumber\\
&+&\cos\theta\bar{\lambda}_1\Gamma^0\lambda_2\bar{\lambda}_4\Gamma_0\lambda_3-\sin\theta\bar{\lambda}_1\Gamma^2\lambda_2\bar{\lambda}_4\Gamma_1\lambda_3-2\frac{\tilde{u}}{\tilde{s}}\bar{\lambda}_1\Gamma^{i_{>1}}\lambda_2\bar{\lambda}_4\Gamma_{i_{>1}}\lambda_3]\, ,
\end{eqnarray}
\begin{eqnarray}
{\cal {T}}_{2,13}
&=& \frac{\tilde{s}\tilde{t}^2}{128}[\bar{\lambda}_1\Gamma^1\lambda_2\bar{\lambda}_4\Gamma_1\lambda_3-\cos\theta\bar{\lambda}_1\Gamma^1\lambda_2\bar{\lambda}_4\Gamma_0\lambda_3-\bar{\lambda}_1\Gamma^0\lambda_2\bar{\lambda}_4\Gamma_1\lambda_3+\cos\theta\bar{\lambda}_1\Gamma^0\lambda_2\bar{\lambda}_4\Gamma_0\lambda_3
\nonumber \\
&-&2\frac{\tilde{u}}{\tilde{s}}\bar{\lambda}_1\Gamma^{i_{>1}}\lambda_2\bar{\lambda}_4\Gamma_{i_{>1}}\lambda_3-\sin\theta\bar{\lambda}_1\Gamma^{2}\lambda_2\bar{\lambda}_4\Gamma_{0}\lambda_3-\sin\theta\bar{\lambda}_1\Gamma^{2}\lambda_2\bar{\lambda}_4\Gamma_{1}\lambda_3] \, ,
\end{eqnarray} 
\begin{eqnarray}
{\cal {T}}_{2,14}
&=&\frac{\tilde{s}\tilde{t}^2}{64}[
 -sin^2\theta\bar{\lambda}_1\Gamma^1\lambda_2\bar{\lambda}_4\Gamma_1 \lambda_3-cos^2\theta\bar{\lambda}_1\Gamma^2\lambda_2\bar{\lambda}_4\Gamma_2 \lambda_3-\bar{\lambda}_1\Gamma^{i_{>2}}\lambda_2\bar{\lambda}_4\Gamma_{i_{>2}} \lambda_3 \nonumber \\
 &+&\cos\theta\bar{\lambda}_1\Gamma^0\Gamma^1\Gamma^{i_{>2}}\lambda_2\bar{\lambda}_4\Gamma_{i_{>2}} \lambda_3+\cos\theta \bar{\lambda}_1\Gamma^0\Gamma^1\Gamma^2\lambda_2\bar{\lambda}_4\Gamma_2 \lambda_3 \nonumber \\
 &+&\sin\theta \bar{\lambda}_1\Gamma^0\Gamma^2\Gamma^{i_{>2}}\lambda_2\bar{\lambda}_4\Gamma_{i_{>2}} \lambda_3-\sin\theta \bar{\lambda}_1\Gamma^0\Gamma^1\Gamma^2\lambda_2\bar{\lambda}_4\Gamma_1 \lambda_3 \nonumber \\
 &+&\sin\theta \cos\theta\bar{\lambda}_1\Gamma^2\lambda_2\bar{\lambda}_4\Gamma_1 \lambda_3+\sin\theta \cos\theta\bar{\lambda}_1\Gamma^1\lambda_2\bar{\lambda}_4\Gamma_2 \lambda_3
] \, ,
\end{eqnarray}
\begin{eqnarray}
{\cal {T}}_{2,15}
&=& -\frac{\tilde{s}\tilde{t}^2}{64}[
\bar{\lambda}_1 \Gamma^{i_{>2}}\lambda_2\bar{\lambda}_4 \Gamma_{i_{>2}} \lambda_3+\bar{\lambda}_1 \Gamma^2\lambda_2\bar{\lambda}_4\Gamma_2 \lambda_3+\bar{\lambda}_1 \Gamma^{i_{>2}}\lambda_2\bar{\lambda}_4 \Gamma_0\Gamma_1\Gamma_{i_{>2}} \lambda_3 \nonumber \\
&+&\bar{\lambda}_1 \Gamma^{2}\lambda_2\bar{\lambda}_4 \Gamma_0\Gamma_1\Gamma_{2} \lambda_3
]\, .
\end{eqnarray}

\newpage

\end{appendices}


\begin{thebibliography}{99}


\bibitem{Kawai:1985xq}
H.~Kawai, D.~C.~Lewellen and S.~H.~H.~Tye,
``A Relation Between Tree Amplitudes of Closed and Open Strings,''
Nucl. Phys. B \textbf{269}, 1-23 (1986)
doi:10.1016/0550-3213(86)90362-7

\bibitem{Martin:2024jpe}
L.~Martin, M.~Parlanti and M.~Schvellinger,
``Fermions hard scattering from superstrings,''
Phys. Rev. D \textbf{111}, no.6, 066004 (2025)
doi:10.1103/PhysRevD.111.066004
[arXiv:2412.08425 [hep-th]].

\bibitem{Veneziano:1968yb}
G.~Veneziano,
``Construction of a crossing - symmetric, Regge behaved amplitude for linearly rising trajectories,''
Nuovo Cim. A \textbf{57}, 190-197 (1968)
doi:10.1007/BF02824451

\bibitem{Virasoro:1969me}
M.~A.~Virasoro,
``Alternative constructions of crossing-symmetric amplitudes with regge behavior,''
Phys. Rev. \textbf{177}, 2309-2311 (1969)
doi:10.1103/PhysRev.177.2309

\bibitem{Shapiro:1970gy}
J.~A.~Shapiro,
``Electrostatic analog for the virasoro model,''
Phys. Lett. B \textbf{33}, 361-362 (1970)
doi:10.1016/0370-2693(70)90255-8

\bibitem{Ramond:1971gb}
P.~Ramond,
``Dual Theory for Free Fermions,''
Phys. Rev. D \textbf{3}, 2415-2418 (1971)
doi:10.1103/PhysRevD.3.2415

\bibitem{Neveu:1971rx}
A.~Neveu and J.~H.~Schwarz,
``Factorizable dual model of pions,''
Nucl. Phys. B \textbf{31}, 86-112 (1971)
doi:10.1016/0550-3213(71)90448-2

\bibitem{Green:1981yb}
M.~B.~Green and J.~H.~Schwarz,
``Supersymmetrical String Theories,''
Phys. Lett. B \textbf{109}, 444-448 (1982)
doi:10.1016/0370-2693(82)91110-8

\bibitem{Green:1981xx}
M.~B.~Green and J.~H.~Schwarz,
``Supersymmetrical Dual String Theory. 2. Vertices and Trees,''
Nucl. Phys. B \textbf{198}, 252-268 (1982)
doi:10.1016/0550-3213(82)90556-9



\bibitem{Thorn:1971jc}
C.~B.~Thorn,
``Embryonic Dual Model for Pions and Fermions,''
Phys. Rev. D \textbf{4}, 1112-1116 (1971)
doi:10.1103/PhysRevD.4.1112

\bibitem{Schwarz:1971uie}
J.~H.~Schwarz,
``Dual quark-gluon model of hadrons,''
Phys. Lett. B \textbf{37}, 315-319 (1971)
doi:10.1016/0370-2693(71)90028-1

\bibitem{Corrigan:1972tg}
E.~Corrigan and D.~I.~Olive,
``Fermion meson vertices in dual theories,''
Nuovo Cim. A \textbf{11}, 749-773 (1972)
doi:10.1007/BF02729477

\bibitem{Brink:1973jd}
L.~Brink, D.~I.~Olive, C.~Rebbi and J.~Scherk,
``The Missing Gauge Conditions for the Dual Fermion Emission Vertex and Their Consequences,''
Phys. Lett. B \textbf{45}, 379-383 (1973)
doi:10.1016/0370-2693(73)90060-9

\bibitem{Olive:1973ewy}
D.~I.~Olive and J.~Scherk,
``Towards satisfactory scattering amplitudes for dual fermions,''
Nucl. Phys. B \textbf{64}, 334-348 (1973)
doi:10.1016/0550-3213(73)90630-5

\bibitem{Mandelstam:1973je}
S.~Mandelstam,
``Manifestly Dual Formulation of the Ramond Model,''
Phys. Lett. B \textbf{46}, 447-451 (1973)
doi:10.1016/0370-2693(73)90163-9

\bibitem{Schwarz:1973jf}
J.~H.~Schwarz and C.~C.~Wu,
``Evaluation of Dual Fermion Amplitudes,''
Phys. Lett. B \textbf{47}, 453-456 (1973)
doi:10.1016/0370-2693(73)90112-3

\bibitem{Schwarz:1974ie}
J.~H.~Schwarz and C.~C.~Wu,
``Functions Occurring in Dual Fermion Amplitudes,''
Nucl. Phys. B \textbf{73}, 77-92 (1974)
doi:10.1016/0550-3213(74)90042-X

\bibitem{Corrigan:1973tye}
E.~Corrigan, P.~Goddard, R.~A.~Smith and D.~I.~Olive,
``Evaluation of the scattering amplitude for four dual fermions,''
Nucl. Phys. B \textbf{67}, 477-491 (1973)
doi:10.1016/0550-3213(73)90210-1

\bibitem{Goddard:1973qh}
P.~Goddard, J.~Goldstone, C.~Rebbi and C.~B.~Thorn,
``Quantum dynamics of a massless relativistic string,''
Nucl. Phys. B \textbf{56}, 109-135 (1973)
doi:10.1016/0550-3213(73)90223-X

\bibitem{Mandelstam:1973jk}
S.~Mandelstam,
``Interacting String Picture of Dual Resonance Models,''
Nucl. Phys. B \textbf{64}, 205-235 (1973)
doi:10.1016/0550-3213(73)90622-6



\bibitem{Schwarz:1982jn}
J.~H.~Schwarz,
``Superstring Theory,''
Phys. Rept. \textbf{89}, 223-322 (1982)
doi:10.1016/0370-1573(82)90087-4

\bibitem{Friedan:1985ge}
D.~Friedan, E.~J.~Martinec and S.~H.~Shenker,
``Conformal invariance, supersymmetry and string theory,''
Nucl. Phys. B \textbf{271}, 93-165 (1986)
doi:10.1016/S0550-3213(86)80006-2

\bibitem{Knizhnik:1985ke}
V.~G.~Knizhnik,
``Covariant Fermionic Vertex in Superstrings,''
Phys. Lett. B \textbf{160}, 403-407 (1985)
doi:10.1016/0370-2693(85)90009-7

\bibitem{Cohn:1986bn}
J.~Cohn, D.~Friedan, Z.~a.~Qiu and S.~H.~Shenker,
``Covariant Quantization of Supersymmetric String Theories: The Spinor Field of the Ramond-neveu-schwarz Model,''
Nucl. Phys. B \textbf{278}, 577-600 (1986)
doi:10.1016/0550-3213(86)90053-2

\bibitem{Green:1987sp}
M.~B.~Green, J.~H.~Schwarz and E.~Witten,
``SUPERSTRING THEORY. VOL. 1: INTRODUCTION,''
1988,
ISBN 978-0-521-35752-4

\bibitem{Knizhnik:1986ke}
V.~G.~Knizhnik,
``Covariant Superstring Fermion Amplitudes From the Sum Over Fermionic Surfaces,''
Phys. Lett. B \textbf{178}, 21-28 (1986)
doi:10.1016/0370-2693(86)90463-6

\bibitem{Gross:1984dd}
D.~J.~Gross, J.~A.~Harvey, E.~J.~Martinec and R.~Rohm,
``The Heterotic String,''
Phys. Rev. Lett. \textbf{54}, 502-505 (1985)
doi:10.1103/PhysRevLett.54.502

\bibitem{Gross:1985fr}
D.~J.~Gross, J.~A.~Harvey, E.~J.~Martinec and R.~Rohm,
``Heterotic String Theory. 1. The Free Heterotic String,''
Nucl. Phys. B \textbf{256}, 253 (1985)
doi:10.1016/0550-3213(85)90394-3

\bibitem{Gross:1985rr}
D.~J.~Gross, J.~A.~Harvey, E.~J.~Martinec and R.~Rohm,
``Heterotic String Theory. 2. The Interacting Heterotic String,''
Nucl. Phys. B \textbf{267}, 75-124 (1986)
doi:10.1016/0550-3213(86)90146-X

\bibitem{Atick:1986rs}
J.~J.~Atick and A.~Sen,
``Covariant One Loop Fermion Emission Amplitudes in Closed String Theories,''
Nucl. Phys. B \textbf{293}, 317-347 (1987)
doi:10.1016/0550-3213(87)90075-7

\bibitem{Becker:2015eia}
K.~Becker, M.~Becker, I.~V.~Melnikov, D.~Robbins and A.~B.~Royston,
``Some tree-level string amplitudes in the NSR formalism,''
JHEP \textbf{12}, 010 (2015)
doi:10.1007/JHEP12(2015)010
[arXiv:1507.02172 [hep-th]].

\bibitem{Green:1980zg}
M.~B.~Green and J.~H.~Schwarz,
``Supersymmetrical Dual String Theory,''
Nucl. Phys. B \textbf{181}, 502-530 (1981)
doi:10.1016/0550-3213(81)90538-1

\bibitem{Green:1981ya}
M.~B.~Green and J.~H.~Schwarz,
``Supersymmetrical Dual String Theory. 3. Loops and Renormalization,''
Nucl. Phys. B \textbf{198}, 441-460 (1982)
doi:10.1016/0550-3213(82)90334-0

\bibitem{Green:1982sw}
M.~B.~Green, J.~H.~Schwarz and L.~Brink,
``N=4 Yang-Mills and N=8 Supergravity as Limits of String Theories,''
Nucl. Phys. B \textbf{198}, 474-492 (1982)
doi:10.1016/0550-3213(82)90336-4

\bibitem{Berkovits:2000fe}
N.~Berkovits,
``Super Poincare covariant quantization of the superstring,''
JHEP \textbf{04}, 018 (2000)
doi:10.1088/1126-6708/2000/04/018
[arXiv:hep-th/0001035 [hep-th]].

\bibitem{Berkovits:2000ph}
N.~Berkovits and B.~C.~Vallilo,
``Consistency of superPoincare covariant superstring tree amplitudes,''
JHEP \textbf{07}, 015 (2000)
doi:10.1088/1126-6708/2000/07/015
[arXiv:hep-th/0004171 [hep-th]].

\bibitem{Alencar:2011tk}
G.~Alencar, M.~O.~Tahim, R.~R.~Landim and R.~N.~Costa Filho,
``RNS and Pure Spinors Equivalence for Type I Tree Level Amplitudes Involving up to Four Fermions,''
[arXiv:1104.1939 [hep-th]].

\bibitem{Berkovits:2022fth}
N.~Berkovits and C.~R.~Mafra,
``Pure Spinor Formulation of the Superstring and Its Applications,''
doi:10.1007/978-981-19-3079-9\_63-1
[arXiv:2210.10510 [hep-th]].

\bibitem{Mafra:2022wml}
C.~R.~Mafra and O.~Schlotterer,
``Tree-level amplitudes from the pure spinor superstring,''
Phys. Rept. \textbf{1020}, 1-162 (2023)
doi:10.1016/j.physrep.2023.04.001
[arXiv:2210.14241 [hep-th]].

\bibitem{Stieberger:2021daa}
S.~Stieberger,
``Open \& Closed vs. Pure Open String One-Loop Amplitudes,''
[arXiv:2105.06888 [hep-th]].

\bibitem{Stieberger:2022lss}
S.~Stieberger,
``A Relation between One-Loop Amplitudes of Closed and Open Strings (One-Loop KLT Relation),''
[arXiv:2212.06816 [hep-th]].

\bibitem{Gross:1986mw}
D.~J.~Gross and J.~H.~Sloan,
``The Quartic Effective Action for the Heterotic String,''
Nucl. Phys. B \textbf{291}, 41-89 (1987)
doi:10.1016/0550-3213(87)90465-2

\bibitem{Garousi:1996ad}
M.~R.~Garousi and R.~C.~Myers,
``Superstring scattering from D-branes,''
Nucl. Phys. B \textbf{475}, 193-224 (1996)
doi:10.1016/0550-3213(96)00316-1
[arXiv:hep-th/9603194 [hep-th]].


\bibitem{Polchinski:2001tt}
J.~Polchinski and M.~J.~Strassler,
``Hard scattering and gauge / string duality,''
Phys. Rev. Lett. \textbf{88}, 031601 (2002)
doi:10.1103/PhysRevLett.88.031601
[arXiv:hep-th/0109174 [hep-th]].

\bibitem{Polchinski:2002jw}
J.~Polchinski and M.~J.~Strassler,
``Deep inelastic scattering and gauge / string duality,''
JHEP \textbf{05}, 012 (2003)
doi:10.1088/1126-6708/2003/05/012
[arXiv:hep-th/0209211 [hep-th]].

\bibitem{Koile:2014vca}
E.~Koile, N.~Kovensky and M.~Schvellinger,
``Hadron structure functions at small $x$ from string theory,''
JHEP \textbf{05}, 001 (2015)
doi:10.1007/JHEP05(2015)001
[arXiv:1412.6509 [hep-th]].

\bibitem{Kovensky:2018xxa}
N.~Kovensky, G.~Michalski and M.~Schvellinger,
``Deep inelastic scattering from polarized spin-$1/2$ hadrons at low $x$ from string theory,''
JHEP \textbf{10}, 084 (2018)
doi:10.1007/JHEP10(2018)084
[arXiv:1807.11540 [hep-th]].

\bibitem{Brower:2006ea}
R.~C.~Brower, J.~Polchinski, M.~J.~Strassler and C.~I.~Tan,
``The Pomeron and gauge/string duality,''
JHEP \textbf{12}, 005 (2007)
doi:10.1088/1126-6708/2007/12/005
[arXiv:hep-th/0603115 [hep-th]].

\bibitem{Brower:2010wf}
R.~C.~Brower, M.~Djuric, I.~Sarcevic and C.~I.~Tan,
``String-Gauge Dual Description of Deep Inelastic Scattering at Small-$x$,''
JHEP \textbf{11}, 051 (2010)
doi:10.1007/JHEP11(2010)051
[arXiv:1007.2259 [hep-ph]].

\bibitem{Jorrin:2022lua}
D.~Jorrin and M.~Schvellinger,
``Scope and limitations of a string theory dual description of the proton structure,''
Phys. Rev. D \textbf{106}, no.6, 066024 (2022)
doi:10.1103/PhysRevD.106.066024
[arXiv:2207.02984 [hep-ph]].

\bibitem{Borsa:2023tqr}
I.~Borsa, D.~Jorrin, R.~Sassot and M.~Schvellinger,
``Proton helicity structure function $g^p_1$ from a holographic Pomeron,''
Phys. Rev. D \textbf{108}, no.5, 056024 (2023)
doi:10.1103/PhysRevD.108.056024
[arXiv:2308.01975 [hep-ph]].



\end{thebibliography}
\end{document}